\documentclass[9pt, a4paper, reqno]{amsart}
\usepackage{mathrsfs}
\usepackage[all]{xy}
\usepackage{geometry, graphicx, amsthm, graphics, amssymb, latexsym}
\usepackage{fancyhdr}
\usepackage{amssymb}
\usepackage{epigraph}
\usepackage{amsmath, amsfonts, amsbsy, pstricks, pst-node, pst-text}
\usepackage{mathrsfs}
\usepackage{amsmath}
\usepackage{multicol}
\usepackage{color}
\theoremstyle{definition}

\begin{document}
\title[Geometrical Foundations of Cartan Gauge Gravity]{Geometrical Foundations of Cartan Gauge Gravity}
\author[Gabriel Catren]{{\sc Gabriel Catren}$^{1,2}$}

\maketitle

\begin{center}
\begin{small}

1- Laboratoire SPHERE (UMR 7219), Universit\'e Paris Diderot - CNRS, Paris,
France.

\end{small}
\end{center}

\begin{center}
\begin{small}

2- Facultad de Filosof\'ia y Letras, Universidad de Buenos Aires - CONICET, Buenos Aires, Argentina.

\end{small}
\end{center}

\date{\today}

\begin{abstract}
We use the theory of \emph{Cartan connections} to analyze the geometrical
structures underpinning the gauge-theoretical descriptions of the gravitational
interaction. According to the theory of Cartan connections, the spin connection
$\omega$ and the soldering form $\theta$ that define the fundamental variables of
the Palatini formulation of general relativity can be understood as different
components of a single field, namely a Cartan connection $A=\omega+\theta$. In
order to stress both the similarities and the differences between the notions
of Ehresmann connection and Cartan connection, we explain in detail how a
Cartan geometry $(P_{H}\rightarrow M, A)$ can be obtained from a $G$-principal
bundle $P_{G}\rightarrow M$ endowed with an Ehresmann connection (being the
Lorentz group $H$ a subgroup of $G$) by means of a bundle reduction mechanism.
We claim that this reduction must be understood as a \emph{partial gauge
fixing} of the local gauge symmetries of $P_{G}$, i.e. as a gauge fixing that
leaves ``unbroken'' the local Lorentz invariance. We then argue that the ``broken''
part of the symmetry--that is the \emph{internal} local translational
invariance--is \emph{implicitly} preserved by the invariance under the
\emph{external} diffeomorphisms of $M$.
\end{abstract}

\begin{small}
\centerline{\em Key words: Gauge Gravity; Ehresmann Connections, Cartan
Connections; Klein Geometries; Atiyah Algebroid.}
\end{small}

\maketitle

\section{Introduction}

The general program aiming to geometrize the fundamental physical interactions
has been developed by means of two kinds of theories. On the one hand, general
relativity describes the gravitational interaction in terms of the \emph{metric
geometry} of spacetime $M$ in such a way that the symmetry group of the theory
is given by the group of diffeomorphisms of $M$. On the other hand, the
Yang-Mills theories describes the non-gravitational interactions in terms of
the \emph{connective geometry} of some ``internal'' spaces over spacetime.
More precisely, the fundamental dynamical structure of a Yang-Mills theory is
an Ehresmann connection on a $G$-principal bundle $P\rightarrow M$ such that
the symmetry group is given by the so-called \emph{gauge group} of vertical
automorphisms of $P$. Since Yang-Mills theories were successfully quantized
(which means in particular that they are renormalizable), a better
comprehension of the relationship between gauge theories and gravitational
theories might provide important hints for the quantization of
gravity.\footnote{By \emph{gauge theory} we do not mean here a
\emph{constrained Hamiltonian system} (since it is a well-known fact that
general relativity can be recast as a Hamiltonian system with constraints),
but rather a theory that describes a dynamical connection on a fiber bundle
over spacetime.} The analysis of this relationship was chiefly developed along
two main avenues of research, namely the attempts to reformulate gravitational
theories as gauge theories (see Refs.\cite{hehl-1976, hehl-1995, kibble,
sciama-1962, trautman-1973, utiyama} and references therein), and the
\emph{gauge theory}/\emph{gravity} dualities such as the AdS/CFT
correspondence. In what follows, we shall address the mathematical underpinning
of a particular strategy for reformulating general relativity as a gauge
theory, namely the strategy based on the theory of \emph{Cartan connections}.

To reformulate general relativity  as a gauge theory means to pass from the
metric Einstein-Hilbert formulation of the theory to a formulation describing a
bundle over spacetime $M$ endowed with a dynamical connection ``gauging'' a
local symmetry. Now, in the framework of the so-called Palatini formulation of
general relativity, this theory can be recast in terms of an Ehresmann
connection $\omega_{H}$ (called \emph{spin connection}) defined in a
$H$-principal bundle over $M$ (where $H$ is the Lorentz group). Roughly speaking, the spin connection is
the gauge field associated to the local Lorentz invariance. However, an
additional field \emph{with no analog in Yang-Mills theories} is also required,
namely the \emph{soldering form} $\theta$. Moreover, the spin connection, far
from being a fundamental structure, can be derived (in the absence of torsion)
from the soldering form. Now, starting in the 70' some people started to consider
the spin connection $\omega_{H}$ and the soldering form $\theta$ as different
components on a single field $A=\omega_{H} + \theta$ \cite{macdowell-mansouri,
stelle-west-1980, Townsend-1977, witten-1988}. In particular, Witten's exact
quantization of (2+1)-dimensional gravity is essentially based on the
utilization of this kind of variables \cite{achucarro-townsend, carlip,
witten-1988}. Now, far from being a mere trick, it can be shown that the
unified entity $\omega_{H} + \theta$ can be understood as a connection. However,
the resulting connection is not an Ehresmann connection, but rather a
\emph{Cartan connection}. From a mathematical viewpoint, the difference between
Yang-Mills theories and general relativity can therefore be understood (at
least at the level of the corresponding geometric structures) in the light of
the difference between Ehresmann and Cartan connections.

In order to understand the presence of the extra field $\theta$ in
gauge-theoretical terms, we could argue as follows. Whereas the spin connection
$\omega$ is the gauge field associated to the \emph{local Lorentz invariance},
we could guess that the soldering form $\theta$ might be understood as the gauge
field associated to some kind of \emph{local translational invariance}. For instance, we could argue (in the case of a vanishing cosmological constant) that $\theta$ might be understood as the gauge field associated to the translational part of the Poincar\'e group. If this were possible, the whole Cartan connection $A=\omega+\theta$ would acquire a straightforward interpretation,
namely that of being the gauge field gauging the whole set of symmetries of Minkowski spacetime. However, the effective implementation of this program is (as we shall see) more convoluted than expected. The reason is that the gauge-theoretical ``internal'' description of the ``external'' geometry of spacetime requires a ``partial gauge fixing'' of the ``internal'' affine symmetries. The gain is that the diffeomorphisms of $M$ acquire--as we shall argue--a gauge-theoretical interpretation.

The main mathematical bibliography on Cartan connections used in this work can
be found in Refs.\cite{alekseevsky-michor, Cap-Slovak-2009, kobayashi-1956,
kobayashi-connections, kobayashi, kobayashi-nomizu, kolar-michor-slovak,
sternberg-rapoport, sharpe} (see also Refs.\cite{Wise, Wise-2009} for a
discussion of Cartan connections in the framework of gravitational physics). In
Section N$^{\circ}$\ref{sect: ehresmann}, we revisit the notion of Ehresmann
connection. In particular, we analyze the localization process associated to
the notion of \emph{local gauge transformations} (or vertical automorphisms of
the corresponding principal bundle) in the light of the theory of Lie
algebroids. In Section N$^{\circ}$\ref{sec: attaching the LHM}, we show that (under a certain condition) a Cartan geometry $(P_{H},A)$ can be obtained from a $G$-principal bundle $P_{G}\rightarrow M$ (with $H$ a subgroup of $G$) endowed with an Ehresmann connection by means of a bundle reduction mechanism. In Section
N$^{\circ}$\ref{sec: cartan geometries}, we discuss the notions of Cartan
connection and Cartan curvature. Section N$^{\circ}$\ref{sec: tetrad} focus on
the resulting geometric interpretation of the soldering form $\theta$. In Section
N$^{\circ}$\ref{sec: diffeos}, we discuss the relation between the
local translational symmetry and the invariance under
diffeomorphisms of spacetime. In Section N$^{\circ}$\ref{sec: tractor}, we show that a Cartan connection $A$ on $P_{H}$
induces an isomorphism between the Atiyah algebroid naturally associated to the bundle
$P_{H}$ and the so-called \emph{adjoint tractor bundle}. This isomorphism permits to
understand the infinitesimal generators of the automorphisms of $P_{H}$ as sections of a
Lie algebra bundle. In Section N$^{\circ}$\ref{sec: transformations},
we compute the transformations of the spin connection $\omega_{H}$, the tetrad $\theta$, the curvature, and the
torsion under both local Lorentz transformations and local gauge translations. In the final Section, we recapitulate the whole construction.

\section{Vertical parallelism and Ehresmann connections}
\label{sect: ehresmann}

In order to analyze the differences and the relations between Ehresmann and Cartan
connections, we shall now revisit the former. Let's
consider a $H$-principal bundle $P\xrightarrow{\pi} M$, where $H$ is a Lie group with Lie algebra $\mathfrak{h}$. The right action $R_{g}: P\rightarrow P$ of $G$ on $P$ induces a Lie algebra homomorphism between the elements in $\mathfrak{h}$ and the corresponding
\emph{fundamental vector fields} on $P$:
\begin{eqnarray}\label{xar}
\mathfrak{h} &\rightarrow& VP\subseteq TP
\\ \xi &\mapsto& X_{\xi}, \nonumber
\end{eqnarray}
where $$X_{\xi}(f(p))=\frac{d}{d\lambda}(f(p \cdot exp(\lambda \xi)))_{|\lambda=0}, \hspace{1 cm} f\in \mathcal{C}^{\infty}(P)$$ and $VP\doteq ker(d\pi)\rightarrow P$ is the (canonical) vertical
subbundle of $TP$. The fundamental vector fields are the infinitesimal generators of the \emph{vertical} $H$-action on $P$. Since, the $H$-action on each fiber is transitive, any vertical vector in $V_{p}P$ can be written as $X_{\xi}(p)$ for some $\xi\in \mathfrak{h}$. Then the map (\ref{xar}) defines a \emph{vertical parallelism}, i.e. a trivialization of the
vertical subbundle $VP$ given by
\begin{eqnarray}\label{yxc}
P \times \mathfrak{h} &\xrightarrow{\simeq}& VP \subseteq TP
\\ (p, \xi) &\mapsto& X_{\xi}(p). \nonumber
\end{eqnarray}

Now, each element $\xi\in \mathfrak{h}$ defines a constant section of the trivial vector bundle $P \times \mathfrak{h}\rightarrow P$. This trivial fact suggests the following natural generalization. Instead of only considering the action of the constant sections of $P \times \mathfrak{h}$ on $P$ (action defined by (\ref{xar})), we can consider the ``local action'' defined by \emph{any} section of $P \times \mathfrak{h}\rightarrow P$. In other terms, we can \emph{localize} the $\mathfrak{h}$-action on $P$ by passing from the constant sections of $P \times \mathfrak{h}\rightarrow P$ to general sections. A section $\sigma\in \Gamma(P \times \mathfrak{h})$ defines (what we shall call) the \emph{localized fundamental vector field} $X_{\sigma}$ on $P$, which is given by
\begin{eqnarray}\label{yrt}
X_{\sigma}(p)=X_{\varphi_{\sigma}(p)}(p),
\end{eqnarray} where $\varphi_{\sigma}: P \rightarrow \mathfrak{h}$ is such that $\sigma(p)=(p, \varphi_{\sigma}(p))$. We can now endow $P \times \mathfrak{h} \rightarrow P$ with a Lie algebroid structure\footnote{A Lie algebroid on $M$ is a vector bundle $\mathcal{A}\rightarrow M$ endowed with 1) a vector bundle map $\Pi: \mathcal{A}\rightarrow TM$ (the so-called \emph{anchor}), and 2) a bracket $[\![\cdot ,\cdot]\!]: \Gamma(\mathcal{A})\times \Gamma(\mathcal{A})\rightarrow \Gamma(\mathcal{A})$ satisfying \begin{eqnarray}\label{dhd}
[\![\sigma, f \varsigma]\!]=f[\![\sigma, \varsigma]\!]+\Pi(\sigma)(f)\varsigma
\end{eqnarray}
and
\begin{eqnarray}\label{bci}
\Pi([\![\sigma, \varsigma]\!])=[\Pi(\sigma),\Pi(\varsigma) ]
\end{eqnarray} for all $\sigma, \varsigma \in \Gamma(\mathcal{A})$, $f\in \mathcal{C}^{\infty}(M)$ \cite{Mackenzie-2005}.} where the anchor is given by (\ref{yxc}) and the bracket between sections is given by the following expression (see Ref.\cite{Mackenzie-2005}, Example 3.3.7, p.104):
\begin{eqnarray}\label{yde}
[\![\sigma_{1}, \sigma_{2} ]\!](p)=(X_{\sigma_{1}}(\sigma_{2}))(p)-(X_{\sigma_{2}}(\sigma_{1}))(p)+(p, [\varphi_{\sigma_{1}}(p),\varphi_{\sigma_{1}}(p)]_{\mathfrak{h}}).
\end{eqnarray}

Now, since $P$ is a $H$-principal bundle over $M$, the $\mathfrak{h}$-action on $P$ can be integrated to the right action of $H$ on $P$. Moreover, the map (\ref{yxc}) is $H$-equivariant (\cite{Mackenzie-2005}, p.93) in the sense that $$(p, \xi)\cdot g \doteq (pg, Ad_{g^{-1}}\xi) \mapsto (R_{g})_{\ast}X_{\xi}(p).$$

Hence, the morphism (\ref{yxc}) of vector bundles over $P$ quotients to the following isomorphism of vector bundles over $M$ (\cite{Mackenzie-2005}, Proposition 3.1.2(ii), p.88): $$P \times_{H} \mathfrak{h} \xrightarrow{\simeq} VP/H,$$
where $P \times_{H} \mathfrak{h}\rightarrow M$ is the called \emph{adjoint bundle}. We can now ``quotient'' the Lie algebroid structure of $P \times \mathfrak{h} \rightarrow P$ by the action of $H$ in order to obtain a Lie algebroid structure in the bundle $P \times_{H} \mathfrak{h}\rightarrow M$. The anchor $P \times_{H} \mathfrak{h}\rightarrow TM$ is obtained by quotiening the anchor (\ref{yxc}) by $H$. Since the $H$-action on $P$ is vertical, the anchor $P \times_{H} \mathfrak{h}\rightarrow TM$ is simply the zero map. The sections $\sigma \in \Gamma(P \times_{H} \mathfrak{h})$ are in bijective correspondence with the $\mathfrak{h}$-valued $H$-equivariant functions on $P$, i.e. with the functions $\varphi_{\sigma}\in \mathcal{C}^{H}(P,\mathfrak{h})$ such that $\varphi_{\sigma}(pg)=Ad_{g^{-1}}(\varphi_{\sigma}(p))$. Indeed, a function $\varphi_{\sigma}\in \mathcal{C}^{H}(P,\mathfrak{h})$ defines a section of $P\times \mathfrak{h}\rightarrow P$ (given by $p\mapsto (p,\varphi_{\sigma}(p))$) that quotients to a section of $P\times_{H} \mathfrak{h}\rightarrow M$ given by $m\mapsto [(p, \varphi_{\sigma}(p))]_{G}$ for any $p$ such that $\pi(p)=m$ (see Appendix N$^{\circ}$\ref{appI}). It is worth noting that the scope of the localization of the $H$-action on $P$ defined by the functions $\varphi_{\sigma}\in \mathcal{C}(P,\mathfrak{h})$ is restricted by passing to the $H$-equivariant functions. The bijection $\Gamma(P \times_{H} \mathfrak{h}) \simeq \mathcal{C}^{H}(P,\mathfrak{h})$ permits us to define a Lie algebra structure $[\![\cdot,\cdot]\!]$ on $\Gamma(P \times_{H} \mathfrak{h})$ by using the Lie bracket in $\mathfrak{h}$ (\cite{Mackenzie-2005}, Proposition 3.2.5, p.95). The Lie bracket $[\![\cdot,\cdot]\!]$ is given by $$[\![\sigma_{1}, \sigma_{2}]\!](m)=[(p,[\varphi_{\sigma_{1}}(p),\varphi_{\sigma_{2}}(p)]_{\mathfrak{h}})]_{G}.$$

Roughly speaking, this bracket results from the bracket (\ref{yde}) by taking the quotient by the $H$-action, eliminating in this way the vertical vector fields in (\ref{yde}). All in all, the fact that the $\mathfrak{h}$-action on $P$ comes from the principal action of $H$ on the bundle $P\rightarrow M$ implies that the action algebroid $P \times \mathfrak{h}\rightarrow P$ quotients to (what we shall call) the \emph{adjoint algebroid} $P \times_{H} \mathfrak{h}\rightarrow M$. In the framework of gauge theories, the sections of $P \times_{H} \mathfrak{h}\rightarrow P$ (or equivalently the functions in $\mathcal{C}^{H}(P,\mathfrak{h}))$ are called \emph{local gauge transformations} and define the so-called \emph{gauge group} of $P$.\footnote{An automorphism $\Phi \in Aut(P)$ of the $H$-principal bundle $P\xrightarrow{\pi} M$ is a diffeomorphism $\Phi: P\rightarrow P$ such that $\Phi(pg)=\Phi(p)g$ for all $p\in P$ and $g\in H$. An automorphism $\Phi$ projects to a unique diffeomorphism $\phi: M \rightarrow M$ such that $\pi\circ \Phi=\phi\circ \pi$. The automorphisms $\Phi$ that project to the identity on $M$ are called \emph{vertical automorphisms} (or \emph{local gauge transformations}) of $P$. The vertical automorphisms define a normal subgroup $Aut_{v}(P)$ of $Aut(P)$ called the \emph{gauge group} of $P$. The infinitesimal generators of the automorphisms of $P_{H}$ are given by the $H$-invariant vector fields on $P_{H}$. This means that the flow $\Phi_{t}$ of an $H$-invariant vector field is such that $\Phi_{t}\in Aut(P_{H})$ for all $t\in \mathbb{R}$. Hence, we shall denote $aut(P)$ the set of $H$-invariant vector fields on $P_{H}$. In turn, $aut_{v}(P)\subset aut(P)$ will denote the set of $H$-invariant \emph{vertical} vector fields on $P_{H}$, i.e. the set of infinitesimal generators of the vertical automorphisms in $Aut_{v}(P)$.}  Roughly speaking, a local gauge transformation $\sigma \in\Gamma(P \times_{H} \mathfrak{h})$ generates an infinitesimal transformation of the frames in $\pi^{-1}(m)\subset P$ defined by $\sigma(m)$ for each $m\in M$. The important point is that the frames in $P$ over different positions in $M$ can be transformed in an independent manner. In the framework of the so-called \emph{gauge-argument}, these local gauge transformations arise by localizing the \emph{global gauge transformations}, i.e. the transformations of the frames in $P$ that do not depend on $m\in M$ (see Ref.\cite{catren2008b} and references therein). As we have just shown, this localization of the global gauge transformations can be understood in terms of the passage from the usual ``global'' $\mathfrak{h}$-action on the $H$-principal bundle $P\rightarrow M$ to the ``localized'' action defined by the sections of the adjoint algebroid $P \times_{H} \mathfrak{h}\rightarrow M$. We could say that the localization process associated to the gauge argument is the physical counterpart of the localization of the global group actions formalized by the notion of Lie algebroid (or Lie groupoid).\footnote{See Ref.\cite{Pradines-2007} for a discussion of this point in the framework of Ehresmann's generalization of Klein's Erlangen program.} The fact that the anchor of the adjoint algebroid is identically zero encodes the fact that the local gauge transformations are purely \emph{internal}, in the sense that they do not relate fibers on different locations in $M$.

An Ehresmann connection on $P\rightarrow M$ is a smooth \emph{horizontal $H$-equivariant
distribution} $\mathcal{H}\subset TP$, i.e. a
subbundle of $TP$ satisfying the following conditions. Firstly, the distribution $\mathcal{H}$ is \emph{horizontal} in the sense that it defines
complementary subspaces to the canonical vertical subspaces:
\begin{eqnarray}\label{neu}
T_{p}P=V_{p}P\oplus \mathcal{H}_{p}.
\end{eqnarray}

Secondly, the horizontal distribution $\mathcal{H}$ is \emph{$H$-equivariant} in the sense that it satisfies the expression
\begin{eqnarray}\label{nbx}
R_{g}^{\ast}\mathcal{H}_{p}=\mathcal{H}_{p g}.
\end{eqnarray}

The \emph{curvature} of $\mathcal{H}$ is defined by means of the following expression (\cite{Cap-Slovak-2009}, p.37)
\begin{eqnarray}\label{xzi}
R(X,Y)=-\Psi([\chi(X), \chi(Y)]), \hspace{1 cm} X,Y
\in \Gamma(TP)
\end{eqnarray}
where the map $\Psi: TP \rightarrow VP$ is the vertical projection with kernel
$\mathcal{H}$, and
\begin{eqnarray}\label{nxd}
\chi= id_{TP}-\Psi
\end{eqnarray}
is the complementary horizontal
projection. By definition, $R$ is horizontal (i.e. it is zero when
at least one of the vectors is a vertical vector) and has vertical values. This definition of the curvature has the advantage of having an immediate intuitive meaning. If $X$ and $Y$ are horizontal vectors, the curvature $R(X,Y)$ is just (minus) the vertical projection of the commutator $[X,Y]$. Therefore, the curvature measures the obstruction preventing the involutivity of the horizontal distribution. If the curvature is zero, the commutator of horizontal vectors is horizontal and the distribution can be integrated.

Given a vector $X \in T_{p}P$, the trivialization (\ref{yxc}) guarantees that there always exists an element $\xi\in \mathfrak{h}$ such that
\begin{eqnarray}\label{nko}
\Psi(X)=X_{\xi}(p).
\end{eqnarray}

Hence, we can define a $\mathfrak{h}$-valued $1$-form on $P$ $$\omega_{G}: T_{p}P\rightarrow \mathfrak{h}$$ given by $$X\mapsto \xi.$$

In particular, $\omega_{p}(X_{\xi}(p))=\xi$. The $H$-equivariance condition (\ref{nbx}) becomes the following condition on the form $\omega_{G}$:
\begin{eqnarray}\label{xcv}
(R_{g}^{\ast}\omega_{pg})(X(p))=\omega_{pg}((R_{g})_{\ast}(X(p))=Ad_{g^{-1}}(\omega_{p}(X(p))),
\end{eqnarray}
or, more succinctly,
\begin{eqnarray}\label{skv}
R_{g}^{\ast}\omega=Ad_{g^{-1}}\circ \omega.
\end{eqnarray}

In other terms, the connection form $\omega$ is a $H$-equivariant map between $TP$ and $\mathfrak{h}$, i.e. the following diagram commutes
\begin{equation*}
\begin{xy}
 \xymatrix{
   T_{p}P \ar[d]_{(R_{g})_{\ast}} \ar[r]^{\omega_{p}}  & \mathfrak{h} \ar[d]^{Ad_{g^{-1}}} \\
   T_{pg}P \ar[r]^{\omega_{pg}} & \mathfrak{h}.
   }
\end{xy}
\end{equation*}

The horizontal distribution $\mathcal{H}$ is given by the kernel of $\omega$:
$$\mathcal{H}_p=\text{Ker } (\omega_{p})\subset T_{p}P.$$

It can be shown that there exists a unique $H$-equivariant horizontal $\mathfrak{h}$-valued $2$-form $F$ on
$P$ such that
$R(X,Y)(p)=X_{F(X(p),Y(p))}(p)$ (\cite{kolar-michor-slovak}, Section 11.2). In terms of $F$ and $\omega$, expression (\ref{xzi})
becomes $$F(X,Y)=-\omega([X-X_{\omega(X)},Y-Y_{\omega(Y)}]).$$

From this expression, it can be deduced the \emph{structure equation}
(\cite{Cap-Slovak-2009}, p.39)\footnote{The space $\Omega(P,\mathfrak{h})$ of $\mathfrak{h}$-valued forms on $P$ can be endowed with the structure of a graded Lie algebra by means of the following bracket (\cite{kolar-michor-slovak}, p.100): $$[\eta, \zeta](X_{1},...,X_{p+q})=\frac{1}{p!q!}\sum sign(\sigma)[\eta(X_{\sigma(1)},...,X_{\sigma(p)}),\zeta(X_{\sigma(p+1)},...,X_{\sigma(p+q)})]_{\mathfrak{h}},$$ where $p$ and $q$ are the degrees of $\eta$ and $\zeta$. In particular, if
$\eta$ and $\zeta$ are $1$-forms, we have $[\eta,\zeta](v,w)=[\eta(v),\zeta(w)]_{\mathfrak{h}}-[\eta(w),\zeta(v)]_{\mathfrak{h}}$.}:

$$F=d\omega+\frac{1}{2}[\omega,\omega].$$

The form $F$ satisfies the so-called \emph{Bianchi identity}: $$d^{\omega}F=0,$$ where $d^{\omega}$ is the so-called \emph{exterior covariant derivative}.\footnote{The \emph{exterior covariant derivative} $d^{\omega}$ is defined by the expression $d^{\omega}\doteq \chi^{\ast}\circ d$, where $\chi$ is the horizontal projection (\ref{nxd}) and $(\chi^{\ast}\eta)(X_{1},..., X_{k})=\eta(\chi(X_{1}),..., \chi(X_{k}))$. It can be shown that the exterior covariant derivative of a $H$-equivariant horizontal $\mathfrak{h}$-valued form $\eta$ is given by the following expression (\cite{kolar-michor-slovak}, p.103):
\begin{eqnarray}\label{nfd}
d^{\omega}\eta=d\eta+[\omega,\eta].
\end{eqnarray}

We can now compute $d^{\omega}F$:
\begin{eqnarray*}
d^{\omega}F &=& dF+[\omega,F]=d(d\omega+\frac{1}{2}[\omega,\omega])+ [\omega,d\omega+\frac{1}{2}[\omega,\omega]]=[d\omega,\omega]+[\omega,d\omega] +\frac{1}{2}[\omega,[\omega,\omega]]=0.
\end{eqnarray*}
where we have used that $[\omega, [\omega,\omega]]=0$ (\cite{sharpe}, Corollary 3.29, p.193).}

We shall now recast the notion of Ehresmann connection in the language of Lie algebroids \cite{Mackenzie-2005}. Let's consider the vector bundle morphism
\begin{equation*}
\begin{xy}
 \xymatrix{
   TP \ar[d] \ar[r]^{T\pi}  & TM \ar[d] \\
   P \ar[r]^{\pi} & M.
   }
\end{xy}
\end{equation*}

Since $\pi(pg)=\pi(p)$ and $T_{pg}\pi\circ T_{p}R_{g}=T_{p}\pi$ (which results from $\pi\circ R_{g}=\pi$), we can take the quotient by $H$ in the bundle on the left thereby obtaining the following diagram (\cite{Mackenzie-2005}, Proposition 3.1.2(i), p.88):
\begin{equation*}
\begin{xy}
 \xymatrix{
   TP/H \ar[d] \ar[r]  & TM \ar[d] \\
   P/H\simeq M \ar[r] & M.
   }
\end{xy}
\end{equation*}

We have thus obtained a morphism $$\Pi: TP/H\rightarrow TM$$ of vector bundles over $M$ (\cite{Mackenzie-2005}, p.92). Clearly, the kernel of this morphism is $VP/H\simeq P\times_{H}\mathfrak{h}$. All in all, we have obtained the so-called \emph{Atiyah exact sequence}
\begin{equation}\label{csq}
0 \rightarrow P\times_{H}\mathfrak{h}\xrightarrow{\iota} TP/H \xrightarrow{\Pi} TM \rightarrow 0,
\end{equation}
of bundles over $M$. Since both $P\times_{H}\mathfrak{h}$ and $TM$ are Lie
algebroids over $M$\footnote{The tangent bundle $TM$ is a trivial Lie algebroid
where the anchor is just the identity map $TM\rightarrow TM$ and the Lie
bracket in $\Gamma(TM)$ is given by the Lie bracket of vector fields on $M$.},
we could try to define a Lie algebroid structure on $TP/H$. This is indeed
possible and the resulting Lie algebroid is the so-called \emph{Atiyah
algebroid} \cite{Atiyah-57}. The anchor of $TP/H$ as a Lie algebroid over $M$
is already given by the map $\Pi$. In order to define a bracket between the
sections of $TP/H \rightarrow M$, we have to use the following isomorphism of
$\mathcal{C}^{\infty}(M)$-modules (\cite{Mackenzie-2005}, Proposition 3.1.4,
p.90)\footnote{Let's consider the following pullback diagram
(\cite{Mackenzie-2005}, proposition 3.1.1,
p.86)
\begin{equation*}
\begin{xy}
 \xymatrix{
   TP \ar[d] \ar[r]^{\natural}  & TP/H \ar[d] \\
   P \ar[r]^{\pi} & M
   }
\end{xy}
\end{equation*}
where the action of $H$ on $TP$ is the differential of the $H$-action on $P$. Then, $\Gamma^{H}(TP)$ is a $\mathcal{C}^{\infty}(M)$-module where $(fX)(p)=(f\circ \pi)(p)X(p)$ for $f\in \mathcal{C}^{\infty}(M)$. The map (\ref{nfe})
is given by $\bar{X}(p)=\natural^{-1}X(\pi(p))$. The inverse $\Gamma^{H}(TP) \xrightarrow{\simeq} \Gamma(TP/H)$ ($\bar{X} \mapsto X$) is given by
$X(m)=\natural (\bar{X} (p)),$ for any $p$ such that $\pi(p)=m$ (if we take $p'=pg$ we have $X(m)=\natural (\bar{X} (pg))=\natural ((R_{g})_{\ast}(\bar{X}(p)))=\natural (\bar{X} (p))$ since $\natural$ is a projection by the $H$-action on $TP$).}
\begin{eqnarray}\label{nfe}
\Gamma(TP/H)&\xrightarrow{\simeq}& \Gamma^{H}(TP)
\\ X &\mapsto& \bar{X}, \nonumber
\end{eqnarray}
where $\Gamma^{H}(TP)$ denotes the $H$-invariant sections of $TP\rightarrow P$, i.e. the vector fields $X$ on $P$ such that $X(pg)=(R_{g})_{\ast}(X(p))$. In other terms, the sections of $TP/H\rightarrow M$ are in bijective correspondence with the $H$-invariant vector fields on $P$. Since $\Gamma^{H}(TP)$ is closed under the Lie bracket of vector fields, we can define a Lie bracket structure $[\![\cdot,\cdot]\!]$ in $\Gamma(TP/H)$ by means of the expression $$\overline{[\![X,Y]\!]}=[\bar{X}, \bar{Y}].$$

It can be shown that this bracket satisfies the expressions (\ref{dhd}) and
(\ref{bci}) (\cite{Mackenzie-2005}, p.94). Therefore, the triplet $(TP/H, \Pi,
[\![\cdot,\cdot]\!])$ defines a Lie algebroid over $M$. The Atiyah exact
sequence (\ref{csq}) encodes the relations between \emph{internal} and
\emph{external} automorphisms of the $H$-principal bundle $P\rightarrow M$.
Firstly, the sections of the trivial algebroid $TM$ generate diffeomorphisms of
$M$, i.e. transformations of the bundle $P\rightarrow M$ which are purely
external in the sense that they only interchange the locations of the different
fibers. Secondly, the sections of the adjoint algebroid
$P\times_{H}\mathfrak{h}$ generate transformations which are purely internal,
that is \emph{vertical} automorphisms of $P$ (or local gauge transformations in
the terminology of physics). Finally, the sections of the Atiyah algebroid
(i.e. the $G$-invariant vector fields on $P$) generate \emph{general}
automorphisms of $P$ (i.e. automorphisms having both internal and external
components). It is worth stressing an important difference between the internal
and the external automorphisms of $P\rightarrow M$: while the former are
obtained by localizing the global action of a Lie algebra $\mathfrak{h}$ on
$P$, this is not the case for the diffeomorphisms of $M$. As we shall see
below, the notion of Cartan connection will allow us to interpret also the
external diffeomorphisms of $M$ in terms of a localized action.

A \emph{Lie algebroid connection} on $(TP/H, \Pi, [\![\cdot,\cdot]\!])$ is a
section of the anchor, i.e. a morphism of vector bundles $\gamma: TM
\rightarrow TP/H$ such that $\Pi\circ \gamma=id_{TM}$ (\cite{Mackenzie-2005},
p.186). Roughly speaking, this definition introduces the notion of connection
by directly defining the horizontal lift of vectors in $TM$. In turn, a
\emph{connection reform} is a morphism of vector bundles $\varpi:
TP/H\rightarrow P\times_{H}\mathfrak{h}$ such that $\varpi\circ
\iota=id_{P\times_{H}\mathfrak{h}}$. It can be shown that there is a bijection
between Lie algebroid connections $\gamma$ and connection forms $\varpi$ such
that (\cite{Mackenzie-2005}, p.187) $$\iota\circ
\varpi+\gamma\circ\Pi=id_{TP/H}.$$

Roughly speaking, this expression can be understood as the $G$-quotient of the
decomposition (\ref{neu}). The curvature of the Lie algebroid connection
measures the deviation of the map $\gamma$ from being a Lie algebra
homomorphism (\cite{Mackenzie-2005}, p.187). More precisely, the curvature
$\bar{R}$ of $\gamma$ is given by the map
\begin{eqnarray}\label{xhg}
\bar{R}: TM\times TM \rightarrow P\times_{H}\mathfrak{h}
\end{eqnarray}
defined by the expression\footnote{It is worth noting that we have two geometric interpretations of the curvature. According to expression (\ref{xzi}), the curvature is the obstruction for the distribution $\mathcal{H}$ to be involutive. According to expression (\ref{xhg}), the curvature is the obstruction for the horizontal lift $\gamma$ to be a Lie algebra homomorphism. The relation between both interpretations can be explained as follows. The commutator $[\gamma (X),\gamma (Y)]$ can be decomposed in its horizontal and vertical components $[\gamma (X),\gamma (Y)]=[\gamma (X),\gamma (Y)]_{h}+[\gamma (X),\gamma (Y)]_{v}$. On the one hand, $[\gamma (X),\gamma (Y)]_{h}=\gamma([X,Y])$ (\cite{kobayashi-nomizu}, proposition 1.3, p.65). On the other hand, $[\gamma (X),\gamma (Y)]_{v}=-R(\gamma (X),\gamma (Y))$. Hence, we have $[\gamma (X),\gamma (Y)]=\gamma([X,Y])-R(\gamma (X),\gamma (Y)),$ which implies $R(\gamma (X),\gamma (Y))=\iota(\bar{R}(X,Y)).$} $$\iota(\bar{R}(X,Y))=\gamma([X,Y])-[\gamma(X),\gamma(Y)].$$

It can be shown that there is a bijective correspondence between Lie algebroid connections on the Atiyah algebroid $(TP/H, \Pi, [\![\cdot,\cdot]\!])$ and Ehresmann connections on $P$ (\cite{Mackenzie-2005}, Proposition 5.3.2, p.193).\footnote{Since a Lie algebroid connection $\gamma$ must be injective (so that $\Pi\circ \gamma=id_{TM}$), its image $\gamma(TM)$ is a vector subbundle of $TP/H$. We can now define a vector subbundle of $TP$ by means of the expression $\mathcal{H}=\natural^{-1}(\gamma(TM))$ where $\natural: TP\rightarrow TP/H$. It is easy to see that $\mathcal{H}$ is a horizontal $H$-equivariance distribution (\cite{Mackenzie-2005}, p.193). Conversely, the decomposition $TP=VP\oplus \mathcal{H}$ quotients to $TP/H=VP/H\oplus \mathcal{H}/H$. Since $\pi_{\ast}: TP/H\rightarrow TM$ is a surjective map with kernel $VP/H$, its restriction to $\mathcal{H}/H$ yields an isomorphism of vector bundles over $M$. The Lie algebroid connection $\gamma: TM \rightarrow \mathcal{H}/H$ is the inverse of $\pi_{\ast}|_{\mathcal{H}/H}$.} Roughly speaking, a section of
the anchor defines $H$-equivariant horizontal subspaces of $TP$, i.e. an Ehresmann connection. In turn, it can be shown that there is a bijection between connection forms $\omega: TP\rightarrow \mathfrak{h}$ and connection reforms $\varpi: TP/H\rightarrow P\times_{H}\mathfrak{h}$ (\cite{Mackenzie-2005}, Proposition 5.3.4, p.194).\footnote{Given the connection reform $\varpi$, the connection form is defined by the expression $\omega(X)= \tau(p,\varpi([X]))$ for $X\in T_{p}P$, where $\tau: P\times(P\times_{H}\mathfrak{h})\rightarrow \mathfrak{h}$ is the obvious map. Conversely, given a connection form view as a map $\omega: TP\rightarrow P\times \mathfrak{h}$, the equivariance condition (\ref{xcv}) permits us to pass to the quotient $\varpi: TP/H\rightarrow P\times_{H} \mathfrak{h}$.}

Let's analyze now the symmetries of the connection form $\omega$. Let's calculate first the Lie
derivative of $\omega$ along a vector field $X
\in \Gamma(TP)$ by using the Cartan's formula $\mathcal{L}_{X}=i_{X}d +di_{X}$:
\begin{eqnarray}\label{vsu}
\mathcal{L}_{X}\omega &=& i_{X}d\omega+d(\omega(X))\nonumber
\\ &=& i_{X}F -\frac{1}{2}([\omega(X), \omega]-[\omega, \omega(X)])+d(\omega(X))\nonumber
\\ &=& i_{X}F +[\omega, \omega(X)]+d\omega(X)
\end{eqnarray}
where $\omega(X)\in \mathcal{C}(P, \mathfrak{h})$. Let's consider in particular a fundamental vector field $X_{\xi}$ obtained by means of the map (\ref{xar}), that is a vertical vector field generating global gauge transformations of $P$. Since the curvature is horizontal and
$\omega(X_{\xi})=\xi$ is a \emph{constant} $\mathfrak{h}$-valued function
on $P_{G}$, we have
\begin{eqnarray}\label{nk}
\mathcal{L}_{X_{\xi}}\omega=[\omega, \xi]=-ad_{\xi}\circ \omega.
\end{eqnarray}

This expression is just the infinitesimal version of the $H$-equivariance condition (\ref{skv}).
Now (as we have argued before), the \emph{global} $\mathfrak{h}$-action on $P$
can be localized by passing from the fundamental vector fields $X_{\xi}$
defined by each $\xi\in \mathfrak{h}$ to the \emph{local} gauge transformations
defined by the sections $\sigma\in\Gamma(P\times_{H}\mathfrak{h})$ of the
adjoint Lie algebroid. A section $\sigma: M \rightarrow
P\times_{H}\mathfrak{h}$ can be naturally lifted to a section $\tilde{\sigma}:
P\rightarrow P\times \mathfrak{h}$.\footnote{The section $\tilde{\sigma}:
P\rightarrow P\times \mathfrak{h}$ is given by $\tilde{\sigma}(p)=
(p,\tau(p,\sigma(\pi(p))))$, where $\tau:P\times
(P\times_{H}\mathfrak{h})\rightarrow \mathfrak{h}$ is the canonical map.} We
can then consider the localized fundamental vector field $X_{\tilde{\sigma}}$
on $P$ defined by expression (\ref{yrt}). We can now calculate the Lie
derivative of $\omega$ under $X_{\tilde{\sigma}}$ by means of expression
(\ref{vsu}). Since the vector field is vertical, the curvature term vanishes.
However, $\omega(X_{\tilde{\sigma}})$ is not necessarily constant. We then have
\begin{eqnarray}\label{yrd}
\mathcal{L}_{X_{\tilde{\sigma}}}\omega &=&[\omega, \omega(X_{\tilde{\sigma}})]+d\omega(X_{\tilde{\sigma}}),
\\ &=& [\omega, \varphi]+d\varphi, \nonumber
\\ &=& d^{\omega}\varphi, \nonumber
\end{eqnarray}
where $\varphi \in \mathcal{C}^{H}(P, \mathfrak{h})=\Omega^{0}_{hor}(P,\mathfrak{h})^{H}$ is given by $\varphi(p)\doteq \omega_{p}(X_{\tilde{\sigma}}(p))$ and we have used expression (\ref{nfd}). We have thus recovered the usual expression for the local gauge transformations of the connection form $\omega$. In what follows, the local gauge transformations defined by $\varphi$ will be denoted $\delta_{\varphi}$. Let's finally calculate the transformation of the curvature under a local gauge transformation:
\begin{eqnarray*}
\delta_{\varphi} F \doteq \mathcal{L}_{X_{\tilde{\sigma}}}F &=& i_{X_{\tilde{\sigma}}}dF+d(F(X_{\tilde{\sigma}}))
\\ &=& i_{X_{\tilde{\sigma}}}(d^{\omega}F -[\omega, F]) \nonumber
\\ &=& -[\omega(X_{\tilde{\sigma}}), F]+[\omega, F(X_{\tilde{\sigma}})] \nonumber
\\ &=& [F, \varphi], \nonumber
\end{eqnarray*}
where we have used the Bianchi identity $d^{\omega}F=0$, and the fact that the curvature is horizontal.

\section{Cartan's program}
\label{sec: attaching the LHM}

As it is nicely explained in Ref.\cite{sharpe}, \emph{Cartan's program} can be
understood as a twofold generalization of both \emph{Riemannian geometry} and
\emph{Klein's Erlangen program}. Briefly, a Cartan
geometry is a manifold which is \emph{infinitesimally modeled by a Klein
geometry} \emph{but locally deformed} by the so-called \emph{Cartan curvature}.
Contrary to Riemannian geometry, the \emph{tangent model} of a Cartan geometry
is not necessarily given by a \emph{flat} model, but rather by a
\emph{maximally symmetric} (or homogeneous) model. Contrary to Klein geometries, the resulting Cartan
geometries are not necessarily maximally symmetric. We could say that whereas Riemannian geometry
stems from the ``localization'' of the Euclidean group, Cartan geometries
result from the ``localization'' of the symmetry group of a Klein
geometry.\footnote{ In particular, a Riemannian geometry on $M$ can be defined
as a torsion-free Cartan geometry on $M$ modeled on Euclidean space (see
Ref.\cite{sharpe}, p.234).} In this way, Cartan's program promotes to the foreground the
gauge-theoretical notion of \emph{local symmetry} to the detriment of Riemann's
notion of \emph{local flatness}: whereas a Riemannian manifold is a
\emph{infinitesimally flat space}, a Cartan geometry is a \emph{infinitesimally
homogeneous space}. Therefore, the local deformation encoded by the Cartan ``curvature'' should not be
understood as a \emph{curving}, but rather as a deviation from the symmetries
of the local Klein models.

We shall now explain how to construct a Cartan geometry on a manifold $M$ by ``attaching'' (zero-order identification) and ``soldering'' (first-order identification) a
Klein geometry to each point of $M$. To do so, let's introduce first the notion of Klein geometry. Let $M_{0}$ be a smooth connected manifold endowed with a \emph{transitive}
action of a Lie group $G$, $m$ a point of $M$, and $H\subset G$ the
isotropy group of $m\in M_{0}$. We shall say that $(M_{0},m)$ is a \emph{Klein
geometry} associated to the pair $(G,H)$ (see Ref.\cite{sharpe}, ch.4). The
group $G$ is called the  \emph{principal group} or the \emph{group of motions}
of the Klein geometry. The map $\pi_{m}:G \rightarrow M_{0}$ given by $g \mapsto g\cdot
m$ is \emph{surjective} (given the transitivity of the action) and induces the following isomorphism
\begin{align*}
G/H\xrightarrow{\simeq} M_{0}
\\ [g] \mapsto g\cdot m.
\end{align*}

In this way, the election of a point $m$ in a $G$-homogeneous manifold $M_{0}$ allows
us to define an isomorphism between $M_{0}$ and $G/H$. In what follows, we shall suppose that the Lie algebra $\mathfrak{g}$ of $G$ is \emph{reductive}, i.e. that there exists a
$ad(H)$-module decomposition of $\mathfrak{g}$, that is a decomposition
\begin{align}\label{ice}
\mathfrak{g}=\mathfrak{h}\oplus \mathfrak{m}
\end{align}
such that $ad(H)\cdot \mathfrak{m}\subset \mathfrak{m}$. In what follows, the
complement $\mathfrak{m}$ to $\mathfrak{h}$ in $\mathfrak{g}$ will be denoted
$\mathfrak{g}/\mathfrak{h}$. In this way, the subalgebra $\mathfrak{h}$ and its complement $\mathfrak{g}/\mathfrak{h}$ satisfy
\begin{align}\label{bwi}
[\mathfrak{h},\mathfrak{h}] &\subseteq \mathfrak{h}
\\ [\mathfrak{h},\mathfrak{g}/\mathfrak{h}] &\subseteq \mathfrak{g}/\mathfrak{h}. \nonumber
\end{align}

We shall also suppose that
$\mathfrak{g}/\mathfrak{h}$ is endowed with an $ad(H)$-invariant scalar product
$\left\langle\cdot, \cdot\right\rangle$.

In order to build a Cartan geometry on $M$, we have to select a pair $(G,H)$ such that $dim(G/H)=dim(M)$. We shall then infinitesimally model the geometry of $M$ by attaching to each $x\in M$ a tangent Klein geometry associated to the pair $(G,H)$.\footnote{Depending on the value of the cosmological constant
$\Lambda$, the homogeneous spaces that are relevant in the framework of
gravitational theories are given by the Minkowski spacetime for $\Lambda=0$, the  de Sitter spacetime for $\Lambda >
0$, and the anti-de Sitter spacetime for $\Lambda< 0$. All these Klein geometries have the same isotropy group, namely the Lorentz group $SO(3,1)$. However, they differ in their group of motions, which are the Poincar\'e group $\mathbb{R}^{4}\rtimes
SO(3,1)$, the de Sitter group $SO(4,1)$, and the anti-de Sitter group $SO(3,2)$ respectively.} To do so, let's introduce a $G$-principal
bundle $P_{G}\xrightarrow{\pi} M$ endowed with an Ehresmann connection
$\omega_{G}$. A fiber in $P_{G}$ is isomorphic to the set of affine frames $(m,e)$ in a Klein geometry associated to the pair $(G,H)$, where an affine frame
consists of: 1) a point $m$ in the Klein geometry and 2) a frame of the tangent space to the Klein geometry at $m$. For the moment, these ``internal'' affine frames have no relation
whatsoever to the tangent spaces to $M$. In fact, we shall show that these
\emph{internal} frames can be
\emph{externalized} so to speak, that is to say \emph{attached} (zero-order
identification) and \emph{soldered} (first-order identification) to $M$. To do so, we shall need additional geometric data, namely 1) a reduction of the $G$-bundle $P_{G}\rightarrow M$ to an $H$-bundle $P_{H}\rightarrow M$ and 2) a soldering form respectively.

The first step for externalizing the internal frames in $P_{G}$ is to \emph{attach} a copy of $G/H$ to each
$x\in M$. This can be done by considering
the associated $G$-bundle in homogeneous spaces $$P_{G}\times_{G} G/H
\xrightarrow{\varrho} M.$$

A point in $P_{G}\times_{G} G/H$ is a $G$-equivalence class of the form $[(p,[g]_{H})]_{G}$. This associated bundle can also be obtained by taking the
quotient of $P_{G}$ by the action of $H$. Indeed, the following bundles are
isomorphic:
$$P_{G}\times _{G} G/H \cong P_{G}/H,$$ where the two
inverse maps are given by $[(p,[g]_{H})]_{G}\mapsto [pg]_{H}$ and
$[p]_{H}\mapsto [(p,[e]_{H})]_{G}$ (see Ref.\cite{kobayashi-nomizu}, Proposition 5.5, p.57). The
fiber $\varrho^{-1}(x)$ will be denoted $M_{0}^{x}$. We shall now select in a smooth manner a
\emph{point of attachment} in $M_{0}^{x}$ for all $x\in M$ by means of a global
section
$$\sigma: M \rightarrow P_{G}\times _{G} G/H \cong P_{G}/H.$$

By doing so, each pair $(M_{0}^{x}, \sigma(x))$ is a Klein geometry associated to the pair $(G,H)$. These Klein geometries  will be called \emph{local homogeneous model} (LHM) of the resulting Cartan geometry. We can then identify $x$ and $\sigma(x)$, that is to say \emph{attach} each LHM $M_{0}^{x}$ to $x$ at the point $\sigma(x)$. It can be shown (see Appendix \ref{appII}) that the data given by the global section $\sigma$ is equivalent to the definition of a $G$-equivariant function:
$$\varphi:P_{G}\rightarrow G/H,  \hspace{1 cm} \varphi(pg)=g^{-1}\varphi_{\sigma}(p).$$

Now, the zero-order attachment provided by the global section $\sigma \in
\Gamma(P_{G}\times _{G} G/H)$ or by the $G$-equivariant function $\varphi \in
\mathcal{C}^{G}(P_{G}, G/H)$ defines a reduction of the original
bundle $P_{G}$ to a $H$-bundle $P_{H}$. The reduced $H$-bundle
$P_{H}$ is given either by the inverse of $[e]_{H}\in G/H$ by $\varphi$ (that
is, $P_{H}=\varphi^{-1}([e]_{H})$) or by the pullback of the projection
$P_{G}\rightarrow P_{G}/H$ along the section $\sigma$ respectively. The reduction process can be summarized by means of the following diagram:

$$
  \xymatrix@R=.8in @C=.55in{
    P_{H}=\varphi_{\sigma}^{-1}([e])=\sigma^* P_{G} \ar@{^{(}->}[r]^{\iota}\ar[d] & P_{G} \ar[d] \ar@{->>}[r]^-{\varphi_{\sigma}} \ar[ld] & G/H  \\
    M  \ar@/^1.2pc/[r]^{\sigma}  &  P_{G}/H \cong P_{G} \times_{G} G/H  \ar[l]                                     &
  }
$$

Whereas a fiber over $x$ in $P_{G}$ is isomorphic to the set of affine frames $(m,e)$ tangent to
$M_{0}^{x}$, a fiber over $x$ in $P_{H}$ is isomorphic to the set of linear frames $(\sigma(x),e)$ in
$T_{\sigma(x)}M_{0}^{x}$. We
could say that the reduction from $P_{G}$ to $P_{H}$ amounts to ``break'' the
affine $G$-symmetry down to the Lorentz $H$-symmetry.\footnote{The (non-canonical) reduction from $P_{G}$ to $P_{H}$ depends on the existence
of a global section $\sigma$ (the obstructions to the existence of
$\sigma$ are analyzed in Ref.\cite{isham}). It is worth noting that, conversely, a bundle $P_{H}$ can always be extended in a canonical manner to a $G$-bundle. The extended $G$-bundle is given by the bundle
$P_{H}\times_{H}G$ associated to $P_{H}$, where the $G$-action is defined by
$[(p,g)]_{H}\cdot g'=[(p,gg')]_{H}$, and where the inclusion $\iota:
P_{H}\hookrightarrow P_{H}\times_{H}G$ is given by $p \mapsto [(p,e)]_{H}$. The important difference between the reduction and the extension is
the latter can \emph{always} be performed \emph{in a
canonical manner}.}

In Ref.\cite{stelle-west-1980}, the reduction process that we have just
described was interpreted in terms of a spontaneous symmetry breaking of the
gauge symmetries of $P_{G}$. In what follows, we shall rather interpret this
reduction as a \emph{partial gauge fixing} of the original $G$-symmetry.
Instead of selecting a single affine frame $(m,e)$ tangent to $M_{0}^{x}$ for
each $x$, the attaching section $\sigma$ just fixes (for each $x$) the point
$m$. In other terms, the section $\sigma$ just selects the $H$-equivalence
class $(\sigma(x),e)$ of linear frames based at the point $m=\sigma(x)$ of
$M_{0}^{x}$. In this way, the section $\sigma$ that defines the reduction can
be understood as a gauge fixing of the translational invariance of each Klein
model. This partial gauge fixing leaves ``unbroken'' the Lorentz gauge
invariance associated to the set of Lorentz frames $(\sigma(x),e)$ for each
$x\in M$. In order to plead in favor of this interpretation of the reduction
process, let's reinterpret an ordinary (i.e. complete) gauge fixing as a
reduction of $P_{G}$ to a $id_{G}$-principal fiber bundle. According to the
reduction process described above, such a reduction is given either by a
section $$\sigma: M \rightarrow P_{G}\times_{G} (G/\left\{id_{G}\right\})\cong
P_{G}/\left\{id_{G}\right\} = P_{G}$$ or by a $G$-equivariant function
$\varphi: P_{G} \rightarrow G/\left\{id_{G}\right\}=G$. The reduced
$\left\{id_{G}\right\}$-bundle is then given by
$P_{\left\{id_{G}\right\}}=\sigma^{\ast}P_{G}=\varphi^{-1}(id_{G})$. From the
whole set of affine frames in a fiber of $P_{G}$, the
$\left\{id_{G}\right\}$-reduction selects a \emph{unique} frame, namely the
frame identified by means of $\varphi$ with the identity in $G$. Therefore, a
gauge fixing can be understood as a complete reduction from $P_{G}$ to
$P_{id_{G}}$. Now, whereas a \emph{gauge fixing} can be interpreted as a
\emph{complete reduction}, a \emph{partial reduction} defined by a non-trivial
subgroup $H$ can be interpreted as a \emph{partial gauge fixing}. Instead of
selecting a unique frame for each $x$, such a partial gauge fixing selects a
non-trivial $H$-class of frames for each $x$.

\section{Absolute parallelism and Cartan Connections}
\label{sec: cartan geometries}

Let's suppose now that the Ehresmann connection $\omega_{G}$ on $P_{G}$
satisfies the following condition:
\begin{align}\label{xus}
Ker(\omega_{G})\cap
\iota_{\ast}TP_{H}=0,
\end{align} where  $\iota: P_{H}\hookrightarrow P_{G}$. In other
terms, we suppose that the restriction $A \doteq \iota^{\ast}(\omega_{G})$ of
$\omega_{G}$ to $P_{H}$ has no null vectors, i.e. that $Ker(A)=0$ (in Section N$^{\circ}$\ref{sec: diffeos} we shall provide a geometric interpretation of this condition). This condition, together with the fact that $dim(G)=dim(P_{H})$,
implies that the $1$-form
\begin{eqnarray}\label{xjr}
A: TP_{H} \rightarrow \mathfrak{g}
\end{eqnarray}
induces a linear isomorphism $T_{p}P_{H} \cong \mathfrak{g}$ for each $p\in
P_{H}$. The $\mathfrak{g}$-valued $1$-form $A$ on $P_{H}$ is called a
\emph{principal Cartan connection} and the pair $(P_{H}, A)$ a \emph{Cartan
geometry on $M$ of type $(G,H)$} (\cite{sharpe}, p.365).
It can be shown that the form $A$ is $H$-equivariant and reproduces the
fundamental vector fields associated to the vertical $H$-action (i.e.
$A_{p}(X_{\xi}(p))=\xi\in \mathfrak{h}$).\footnote{It is worth noting that a
principal Cartan connection on a $H$-principal bundle $P_{H}\rightarrow M$ can
be directly defined without passing by the bundle reduction of a $G$-bundle
$P_{G}\rightarrow M$. To do so, we do not need a Lie group $G$, but just--what
Sharpe calls--a \emph{model geometry}, that is 1) a pair $(\mathfrak{g}, H)$
and 2) a representation $Ad$ of $H$ on $\mathfrak{g}$ extending the adjoint
representation of $H$ on $\mathfrak{h}$ (\cite{sharpe}, p.174).}

It is worth stressing that the form $A$ cannot be understood as an Ehresmann
connection on $P_{H}$. Indeed, the form $A$ is not valued in the Lie algebra
$\mathfrak{h}$ of the structural group $H$ of $P_{H}$, but rather in the bigger
Lie algebra $\mathfrak{g}$. We can understand this fact by remarking that a
principal Cartan connection extends the natural \emph{$\mathfrak{h}$-valued
vertical parallelism} $P_{H} \times \mathfrak{h} \xrightarrow{\simeq}  VP_{H}$
given by (\ref{yxc}) to a \emph{$\mathfrak{g}$-valued absolute parallelism}
given by (\cite{Cap-Slovak-2009}, p.71)
\begin{eqnarray}\label{xuy}
P_{H} \times \mathfrak{g} &\xrightarrow{\simeq}& TP_{H}
\\ (p, \xi) &\mapsto& X_{\xi}(p)=A_{p}^{-1}(\xi). \nonumber
\end{eqnarray}

The vector fields on $P_{H}$ obtained by means of
the map \begin{eqnarray*}
\zeta^{A}: \mathfrak{g} &\rightarrow& TP_{H}
\\ \xi &\rightarrow& X_{\xi}
\end{eqnarray*} are called \emph{parallel vector fields}. While the vertical
parallelism $P_{H} \times \mathfrak{h} \xrightarrow{\simeq}  VP_{H}$ is
\emph{naturally} defined by the vertical action of the structural group $H$ on
$P_{H}$, the $\mathfrak{g}$-valued absolute parallelism (\ref{xuy}) depends on
the principal Cartan connection $A$.

By composing with the projections $\pi_{\mathfrak{h}}:
\mathfrak{g} \rightarrow \mathfrak{h}$ and $\pi_{\mathfrak{g}/\mathfrak{h}}:
\mathfrak{g} \rightarrow \mathfrak{g}/\mathfrak{h}$, the
$H$-invariant decomposition
$\mathfrak{g}=\mathfrak{h}\oplus \mathfrak{g}/\mathfrak{h}$ induces the
following decomposition of the principal Cartan connection $A$:
$$A=\omega_{H}+\theta,$$ where
$$\omega_{H}: TP_{H} \xrightarrow{A} \mathfrak{g} \xrightarrow{\pi_{\mathfrak{h}}} \mathfrak{h}$$ and $$\theta:TP_{H} \xrightarrow{A} \mathfrak{g} \xrightarrow{\pi_{\mathfrak{g}/\mathfrak{h}}} \mathfrak{g}/ \mathfrak{h}.$$

On the one hand, it can be shown that $\omega_{H}\in \Omega^{1}(P_{H},
\mathfrak{h})$ is an Ehresmann connection on $P_{H}$. In the physics
literature (where $H$ is the Lorentz group), the Ehresmann connection $\omega_{H}$ is called
\emph{spin connection}. On the other hand, it can be shown that the so-called
\emph{soldering form}
 $\theta$ is a \emph{horizontal} (i.e., $\theta(\eta)=0$ for vertical vectors $\eta\in VP_{H}$) and \emph{$H$-equivariant} (i.e., $R^{\ast}_{h}\theta=h^{-1}\theta$) $\mathfrak{g}/ \mathfrak{h}$-valued $1$-form on $P_{H}$.

The \emph{Cartan curvature} $F\in \Omega^{2}(P_{H},\mathfrak{g})$ of a Cartan
geometry $(P_{H}, A)$ is given by the structure equation
$$F=dA +\frac{1}{2}[A,A]=F_{\mathfrak{h}}+F_{\mathfrak{g}/\mathfrak{h}},$$
where $F_{\mathfrak{h}}\doteq \pi_{\mathfrak{h}}\circ F$ and
$F_{\mathfrak{g}/\mathfrak{h}}\doteq \pi_{\mathfrak{g}/\mathfrak{h}}\circ F$.
It can be shown that $F$ is a horizontal form (\cite{Cap-Slovak-2009}, Lemma
1.5.1, p.72). In turn, the \emph{curvature} $R\in
\Omega^{2}(P_{H},\mathfrak{h})$ and the \emph{torsion} $T\in
\Omega^{2}(P_{H},\mathfrak{g}/\mathfrak{h})$ of a Cartan geometry $(P_{H}, A)$
are defined by the expressions
$$R \doteq   d\omega_{H}+\frac{1}{2}[\omega_{H},\omega_{H}]=F_{\mathfrak{h}}-\frac{1}{2}[\theta,\theta]_{\mathfrak{h}}$$
and $$T \doteq   d\theta+[\omega_{H},\theta]=F_{\mathfrak{g}/\mathfrak{h}}-\frac{1}{2}[\theta,\theta]_{\mathfrak{g}/\mathfrak{h}}.$$

The Cartan curvature satisfies the usual Bianchi identity $$d^{A}F=0.$$

From this identity we can derive the Bianchi identities for the curvature $R$ and the torsion $T$\footnote{Indeed,
\begin{eqnarray*}
d^{A}F &=& dF+[A,F]
\\ &=& d(R+T+\frac{1}{2}[\theta,\theta])+[\omega+\theta, R+T+\frac{1}{2}[\theta,\theta]]
\\ &=& dR+dT+\frac{1}{2}d[\theta,\theta]+[\omega, R] + [\omega, T] + \frac{1}{2}[\omega, [\theta,\theta]]+[\theta, R] + [\theta, T] + \frac{1}{2}[\theta, [\theta,\theta]]
\\ &=& d^{\omega}R+ d^{\omega}T+[\theta, R]+ [d\theta,\theta]+\frac{1}{2}[\omega, [\theta,\theta]]+ [\theta, d\theta + [\omega,\theta]]
\\ &=& d^{\omega}R+ d^{\omega}T+[\theta, R]+ [d\theta,\theta]+\frac{1}{2}[\omega, [\theta,\theta]]+ [\theta, d\theta ]+[\theta, [\omega,\theta]]
\\ &=& d^{\omega}R+ d^{\omega}T+[\theta, R]+\frac{1}{2}2[[\omega,\theta], \theta]+[\theta, [\omega,\theta]]
\\ &=& d^{\omega}R+ d^{\omega}T+[\theta, R],
\end{eqnarray*}
where we have used that $[\theta, [\theta,\theta]]=0$ (\cite{sharpe}, Corollary 3.29, p.193) and that $[\omega, [\theta,\theta]]=2 [[\omega,\theta], \theta]$ (\cite{sharpe}, Corollary 3.28, p.192).
}:
\begin{eqnarray*}
d^{\omega_{H}}R &=& 0
\\ d^{\omega_{H}}T &=& [R,\theta].
\end{eqnarray*}

Let's analyze the meaning of the notion of Cartan curvature. By contracting the
Cartan curvature with two parallel vector fields we obtain the
expression\footnote{Indeed,
\begin{eqnarray*}
F(X_{\xi},X_{\eta})&=&(dA)(X_{\xi},X_{\eta})+[A(X_{\xi}), A(X_{\eta})]_{\mathfrak{g}}
\\ &=&X_{\xi}A(X_{\eta})-X_{\eta}A(X_{\xi})-A([X_{\xi},X_{\eta}]_{\Gamma(TP)})+[A(X_{\xi}), A(X_{\eta})]_{\mathfrak{g}}
\\ &=&-A([X_{\xi},X_{\eta}]_{\Gamma(TP)})+[A(X_{\xi}), A(X_{\eta})]_{\mathfrak{g}}.
\end{eqnarray*}}

$$F(X_{\xi},X_{\eta})=-A([X_{\xi},X_{\eta}]_{\Gamma(TP)})+[A(X_{\xi}),
A(X_{\eta})]_{\mathfrak{g}}.$$

In turn, this entails the expression
\begin{eqnarray}\label{cjk}
\zeta^{A}(F(X_{\xi},X_{\eta}))&=&-[X_{\xi},X_{\eta}]_{\Gamma(TP)}+\zeta^{A}([A(X_{\xi}), A(X_{\eta})]_{\mathfrak{g}})
\\ &=&-[\zeta^{A}(\xi),\zeta^{A}(\eta)]_{\Gamma(TP)}+\zeta^{A}([\xi, \eta]_{\mathfrak{g}}). \nonumber
\end{eqnarray}

Therefore, the Cartan curvature measures the extent to which the
$\mathfrak{g}$-valued absolute parallelism on $P_{H}$ defined by the principal
Cartan connection fails to be a Lie algebra homomorphism between $\mathfrak{g}$
and $\Gamma(TP)$ (\cite{alekseevsky-michor}, 2.7).

It is worth stressing that in general \emph{Cartan flatness} $F=0$ does not
imply $R=0$ and $T=0$:
\begin{align}\label{xor}
F=0 \Leftrightarrow \left\{
\begin{array}{ll}
R_{0}=-\frac{1}{2}[\theta,\theta]_{\mathfrak{h}}\\
T_{0}=-\frac{1}{2}[\theta,\theta]_{\mathfrak{g}/\mathfrak{h}}
\end{array} \right.
\end{align}

The so-called \emph{symmetric models} (like for instance Minkoswki, de Sitter,
and anti-de Sitter spacetime) satisfy--in addition to (\ref{bwi})--the
expression:
$$[\mathfrak{g}/\mathfrak{h},\mathfrak{g}/\mathfrak{h}]\subseteq \mathfrak{h}.$$

For the symmetric models, $T=F_{\mathfrak{g}/\mathfrak{h}}$ and therefore
$$F=(R+\frac{1}{2}[\theta,\theta]_{\mathfrak{h}})+T,$$ which means that the
torsion $T$ naturally appears as the translational component of $F$. If the
model satisfies in addition the stronger condition
$[\mathfrak{g}/\mathfrak{h},\mathfrak{g}/\mathfrak{h}]=0$, then the Cartan
curvature is just the sum of the standard curvature and the torsion: $F=R+T$.
In any case, the curvature and the torsion can be understood as different
components of a unique Cartan curvature. In the so-called Einstein-Cartan
theory of gravity, the torsion, far from being constrained to be zero as in
general relativity, depends on the spin density \cite{hehl-1976}. Therefore,
the Einstein-Cartan theory coincides with general relativity in the absence of
spinor fields. If $T=0$, the spin connection $\omega_{H}$ is completely fixed by $\theta$.
However, $\omega_{H}$ and $\theta$ are independent geometric structures in the
general case. This means that in general the notion of parallelism defined by
$\omega_{H}$ is decoupled from the geometric structures defined by $\theta$ (that is, as we shall see, a \emph{metric} on $M$ and the notion, proposed by Cartan, of \emph{development}). The elegance of Cartan's formalism is that the geometric structures defined by $\omega_{H}$ and $\theta$ are unified into the unique geometric
structure defined by the Cartan connection $A=\omega_{H} + \theta$.

A canonical example of a Cartan connection is provided by the
\emph{Maurer-Cartan form} $A_{G}$ of a Lie group $G$.\footnote{The
\emph{Maurer-Cartan form} $A_{G}(g):T_{g}G \rightarrow \mathfrak{g}$ is defined
by $\xi \mapsto (L_{g^{-1}})_{\ast}\xi$, where $L_{g^{-1}}:G\rightarrow G$ is
the left translation defined by $L_{g^{-1}}(a)=g^{-1}a$ \cite{sharpe}.} Indeed,
$A_{G}$ is a $\mathfrak{g}$-valued $1$-form on the total space $G$ of the
canonical $H$-fibration $G\rightarrow G/H$ defined by the pair $(G,H)$. The
so-called \emph{Maurer-Cartan structure equation}
\begin{align*}
F=dA_{G}+\frac{1}{2}[A_{G},A_{G}]=0
\end{align*}
means that the Cartan connection $A_{G}$ on $G\rightarrow G/H$ is Cartan flat.
This example shows that the standard for Cartan flatness $F=0$ is given
by the Klein geometry $(G/H, [id_{G}])$ canonically associated to the pair $(G,H)$. Now,
this Klein geometry does not necessarily satisfy the flatness and
torsion-free conditions $R=0$ and $T=0$. We could then say that the Cartan
curvature does not measure the deficiency of standard flatness
and torsionfreeness, but rather the deficiency
of symmetry. In this way, a Cartan geometry $(P_{H},A)$ on $M$ of type $(G,H)$
is the non-homogeneous generalization of the Cartan geometry $(G,A_{G})$ defined by the Klein Geometry $(G/H, [id_{G}])$ endowed with the canonical (Maurer-)Cartan connection $A_{G}$.

The example provided by the Cartan geometry $(G,A_{G})$ motivates the
following definitions. The principal idea is that we can also understand a
principal Cartan connection as a \emph{deformation} of a
\emph{$\mathfrak{g}$-structure} on $P_{H}$. By following
Ref.\cite{alekseevsky-michor}, we shall say that a manifold $P$ has a
\emph{$\mathfrak{g}$-structure} (with $dim(P)=dim(\mathfrak{g})$) if it is endowed with a $\mathfrak{g}$-valued
$1$-form $\kappa$ which 1) is non-degenerate in the sense that
$\kappa_{p}:T_{p}P\rightarrow \mathfrak{g}$ is a linear isomorphism for each
$p\in P$ and 2) satisfies the Maurer-Cartan structure equation
$$d\kappa+\frac{1}{2}[\kappa,\kappa]=0.$$

Given a free and \emph{transitive} action of a Lie algebra $\mathfrak{g}$ on a
manifold $P$, there is a $\mathfrak{g}$-structure on $P$ given by
$\kappa_{p}=\zeta_{p}^{-1}$, where $\zeta: \mathfrak{g}\rightarrow \Gamma(TP)$
is the natural map between $\mathfrak{g}$ and the fundamental vector fields on
$P$ (\cite{alekseevsky-michor-1995}, Section N$^{\circ}$5, p.17). A manifold
$P$ endowed with a free and transitive action of a Lie group $G$ is isomorphic
to $G$. Indeed, each point $p\in P$ defines an isomorphism
$\pi_{p}:G\xrightarrow{\simeq} P$ given by $g \mapsto g\cdot p$. Now, if $P$ is
just endowed with a $\mathfrak{g}$-structure, there exists for each $p\in P$ a
unique $\mathfrak{g}$-equivariant local diffeomorphism $C_{p}:P\rightarrow
W_{p}$ such that $C_{p}(p)=id_{G}$, where $W_{p}$ is a connected open set in
$G$ (\cite{alekseevsky-michor-1995}, Section N$^{\circ}$5.2). According to Ref.\cite{alekseevsky-michor}, a \emph{Cartan connection of
type $\mathfrak{g}/\mathfrak{h}$} on a manifold $P$ of dimension
$n=dim(\mathfrak{g})$ is a $\mathfrak{g}$-valued $1$-form on $P$
$$A:TP\rightarrow \mathfrak{g}$$ defining an isomorphism $T_{p}P \simeq \mathfrak{g}$ for each $p\in �$ and such that
\begin{eqnarray}\label{xsz}
[X_{\xi},X_{\eta}]=X_{[\xi,\eta]}
\end{eqnarray}
for $\xi\in \mathfrak{g}$ and $\eta\in \mathfrak{h}$. This means that we have a
free action of $\mathfrak{h}$ on $P$ given by $\zeta^{A}|_{\mathfrak{h}}$. When
this $\mathfrak{h}$-action integrates to a free and proper action of a Lie
group $H$, the orbit space $M\doteq P/H$ is a smooth manifold and the Cartan
connection of type $\mathfrak{g}/\mathfrak{h}$ defines a principal Cartan
connection on $P\rightarrow M$ (\cite{alekseevsky-michor}, p.2). In this
framework, the manifold $M$, far from being presupposed, arises as the orbit
space obtained by integrating the ``vertical'' $\mathfrak{h}$-action on $P$. If the principal Cartan connection is
flat (i.e. if $A$ satisfies the Maurer-Cartan structure equation), expression
(\ref{cjk}) implies that (\ref{xsz}) holds for every $\xi,\eta \in
\mathfrak{g}$. Therefore, a \emph{flat} principal Cartan connection on a
$H$-principal bundle $P_{H}\rightarrow M$ defines an action of the whole Lie algebra
$\mathfrak{g}$ on $P_{H}$, i.e. a $\mathfrak{g}$-structure on $P_{H}$. In
general, we can understand a \emph{curved} principal Cartan connection on
$P_{H}$ as a \emph{deformation} of a $\mathfrak{g}$-structure, in the sense
that expression (\ref{xsz}) does not necessarily holds for every $\xi,\eta \in
\mathfrak{g}$. While the subalgebra $\mathfrak{h}\subset \mathfrak{g}$
necessarily defines a (vertical) Lie algebra action on $P_{H}$, the curvature
of the Cartan connection encodes the obstruction to the extension of this
action to the whole Lie algebra $\mathfrak{g}$.

\section{On the Soldering Form}
\label{sec: tetrad}

As we have explained in Section N$^{\circ}$\ref{sec: attaching the LHM}, the
global section $\sigma$ \emph{attaches} a LHM $M_{0}^{x}$ to each $x\in M$ at
the \emph{point of attachment} $\sigma(x)$ of $M_{0}^{x}$. We shall now show
that the soldering form $\theta$ defined by the principal Cartan connection $A$
enriches the zero-order identification defined by $\sigma$ by identifying each
$T_{x}M$ with $T_{\sigma(x)}M_{0}^{x}$. In this way, the LHM $M_{0}^{x}$
attached to $x$ by means of $\sigma$ will also be \emph{soldered} to $M$ at the point of attachment $\sigma(x)$.

Given a soldering form $$\theta:TP_{H} \rightarrow \mathfrak{g}/\mathfrak{h},$$
we can define the following $1$-form on $M$: $$\tilde{\theta}: TM \rightarrow
E\doteq P_{H} \times_{H}\mathfrak{g}/\mathfrak{h}.$$ The relation between
these two forms results from the following isomorphism:
\begin{align}\label{ynp}
\Omega^{q}_{hor}(P_{H},\mathfrak{g}/\mathfrak{h})^{H}\simeq \Omega^{q}(M,E),
\end{align}
where $\Omega^{q}_{hor}(\cdot,\cdot)^{H}$ denotes the horizontal and
$H$-equivariant differential $q$-forms. The bundle $P_{H} \times_{H}\mathfrak{g}/\mathfrak{h}$ can be
identified with the bundle of vectors tangent to the fibers of
$P_{G}\times_{G}G/H$ along the section $\sigma$ (\cite{sternberg-rapoport},
p.452): $$P_{H} \times_{H}\mathfrak{g}/\mathfrak{h} \simeq V_{\sigma}(P_{G}\times _{G} G/H).$$

Indeed, the fiber at $x$ of the $H$-principal bundle $P_{H}$ obtained by
reducing $P_{G}$ by means of the section $\sigma: M\rightarrow
P_{G}\times_{G}G/H$ is composed of frames tangent to
$M_{0}^{x}\simeq G/H$ at $\sigma(x)$. The vectors in the associated vector bundle $P_{H} \times_{H}\mathfrak{g}/\mathfrak{h}$ are vectors framed by the frames in $P_{H}$ with coordinates in $\mathfrak{g}/\mathfrak{h}$, i.e. vectors tangent to the fibers $M_{0}^{x}$ of $P_{G}\times_{G}G/H$ at $\sigma(x)$.

On the one hand, the $1$-form
$\tilde{\theta}: TM \rightarrow E$ establishes an identification between
vectors $\tilde{v}\in TM$ and vectors in $E$ \emph{in a coordinate-independent
manner}. Therefore, the form $\tilde{\theta}$ will be called from now
\emph{geometric} soldering form. On the other hand, the associated form $\theta$
evaluated at a vector $v\in TP_{H}$ gives the coordinates of the vector
$\tilde{\theta}(\pi_{\ast}v)\in E$ in the ``frame'' $p$.\footnote{Indeed,
$\theta_{p}(v)=\tau(p,\tilde{\theta}(\pi_{\ast}v))$, where we have used
expression (\ref{xiz}) in the Appendix \ref{appII}.} In this way, $\theta$
identifies vectors in $TM$ with vectors in $E$ by encoding the coordinates of
the later in all possible ``frames'' $p\in P_{H}$. Therefore, the form $\theta$
will be called from now on \emph{coordinate} soldering form. Whereas the section $\sigma$ induces a zero-order identification between $M$
and the LHM $M_{0}^{x}$ by identifying each $x\in M$ with the point of attachment
$\sigma(x)$ of $M_{0}^{x}$, the geometric soldering form $\tilde{\theta}: TM
\rightarrow E$ identifies each tangent space $T_{x}M$ with the tangent space to
the internal LHM $M_{0}^{x}$ at $\sigma(x)$. Therefore, the geometric data defined by the attaching section $\sigma$ and the soldering form $\tilde{\theta}$ amount to
\emph{attach} to each $x$ a \emph{tangent} LHM $M_{0}^{x}$. In this way, we
have accomplished Cartan's program, that is, we have substituted the local
\emph{flat} models of Riemannian geometry by the local \emph{homogeneous} models
given by the Klein geometries $(M_{0}^{x}, \sigma(x))\simeq
G/H$.

We shall now explain how the soldering form induces a metric on $M$ by using the $ad(H)$-invariant metric on $\mathfrak{g}/\mathfrak{h}$. It can be shown that the soldering
form defines an injective $H$-morphism
\begin{eqnarray}\label{het}
f^{\theta}: P_{H}\hookrightarrow \mathcal{F}M
\end{eqnarray}
between each $p \in P_{H}$ and a frame $f^{\theta}(p) :
\mathfrak{g}/\mathfrak{h} \rightarrow T_{\pi(p)}M$ in the frame bundle
$\mathcal{F}M$ \cite{sternberg-1985}.\footnote{The frame bundle $\mathcal{F}M
\rightarrow M$ can be defined as follows. Given a $n$-dimensional vector space
$(V,e)$ with a distinguished basis $e$, we can define a frame on $T_{x}M$ as an
isomorphism $f: V\rightarrow T_{x}M$. Given a frame $f$, any $v\in T_{x}M$ can
be expressed as $f(v)$ for some $v\in V$, i.e. as a pair $(f,v)$. We can now
define a right action of $Gl(V)$ on the frames, where $f\cdot g:V\rightarrow
T_{x}M$ is given by $(f\cdot g)(v)= f(g\cdot v)$. In turn, $Gl(V)$ acts on the
pairs $(f,v)$ by means of the expression $(f,v)\cdot g=(f\cdot g, g^{-1}\cdot
v)$. This action guarantees that the pairs $(f,v)$ and $(f,v)\cdot g$ yield the
same vector in $T_{x}M$ (indeed, $(f\cdot g)(g^{-1}\cdot v)=f(v)\in T_{x}M$).
Hence, any vector $f(v)$ in $T_{x}M$ can be identified with the class
$[(f,v)]_{Gl(V)}$. We have thus shown that $TM \simeq
\mathcal{F}M\times_{Gl(V)}V$.} This morphism is given by
$$f^{\theta}(p)(\xi) = \tilde{\theta}_{p}^{-1}(x)(\xi), \hspace{1 cm} x=\pi(p), \hspace{0.5 cm} \xi\in \mathfrak{g}/\mathfrak{h}$$ where
\begin{eqnarray}\label{nbv}
\tilde{\theta}_{p}(x):T_{x}M \xrightarrow{\tilde{\theta}(x)} E
\xrightarrow{\tau(p,\cdot)} \mathfrak{g}/\mathfrak{h}
\end{eqnarray} and $\tau:
P_{H}\times_{M}E \rightarrow \mathfrak{g}/\mathfrak{h}$ is the canonical map
(see Appendix \ref{appII}).\footnote{Let's show that (\ref{het}) is indeed an
$H$-morphism. We want to see that $f^{\theta}(ph)(\xi)=(f^{\theta}(p)\cdot
h)(\xi)=f^{\theta}(p)(h\cdot \xi)$. Let's evaluate the frame
$f^{\theta}(ph):\mathfrak{g}/\mathfrak{h}\rightarrow T_{x}M$ at a vector
$\xi\in \mathfrak{g}/\mathfrak{h}$ by using the inverse of (\ref{nbv}). The
inverse of $\tau(ph,\cdot)$ applied to $\xi$ yields $[(ph, \xi)]\in E$. Now,
this element can also be represented by means of the pair $(p,h\xi)$.} This
application identifies each element $p$ in $P_{H}$ with a frame
$f^{\theta}(p)\in \mathcal{F}M$ over $\pi(p)\in M$. Since $f^{\theta}:
P_{H}\rightarrow \mathcal{F}M$ is an injective $H$-morphism,
$f^{\theta}(P_{H})$ is a $H$-subbundle of the
$GL(\mathfrak{g}/\mathfrak{h})$-frame bundle $\mathcal{F}M$. Now, the reduction
of the $GL(\mathfrak{g}/\mathfrak{h})$-frame bundle $\mathcal{F}M$ to an
$H$-bundle allows to use the $Ad(H)$-invariant scalar product in
$\mathfrak{g}/\mathfrak{h}$ to define a scalar product on the tangent spaces
$T_{x}M$.\footnote{Let's fix a scalar product $\langle\cdot,\cdot\rangle_{V}$
in $V$ such that the privileged basis $e$ is an orthonormal basis. We can now
calculate the scalar product of two vectors in $T_{x}M$ by means of the
expression $\langle (f,v), (f,v')\rangle_{T_{x}M}\doteq \langle v,
v'\rangle_{V}$. Now, in order to be a genuine scalar product in $T_{x}M$,
$\langle \cdot, \cdot\rangle_{T_{x}M}$ should not depend on the chosen
representatives in the classes $[(f,v)]$ and $[(f,v')]$. This is the case
if we restrict the available frames in $\mathcal{F}M$ to the frames that are
related by elements in $O(n)$. In that case, $\langle (f\cdot g,g^{-1}\cdot v),
(f\cdot g,g^{-1}\cdot v')\rangle_{T_{x}M}\doteq \langle g^{-1}\cdot v,
g^{-1}\cdot v'\rangle_{V}=\langle v, v'\rangle_{V}$ since
$\langle\cdot,\cdot\rangle_{V}$ is $O(n)$-invariant. In other terms, the scalar
product $\langle\cdot,\cdot\rangle_{V}$ induces a scalar product in each
tangent space if we reduce the frame bundle $\mathcal{F}M \rightarrow M$ to a
$O(n)$-bundle.} Therefore, the $H$-morphism $f^{\theta}$ induced by the
soldering form $\theta$ defines a metric $g^{\theta}$ on $M$
\cite{kobayashi-nomizu, sternberg-1964}. The metric $g^{\theta}$ on $M$ can be
explicitly defined in terms of the scalar product $\left\langle\cdot,
\cdot\right\rangle_{\mathfrak{g}/\mathfrak{h}}$ by means of the expression
$$g^{\theta}(v,w)=\left\langle\tilde{\theta}_{p}(x)(v),\tilde{\theta}_{p}(x)(w)\right\rangle_{\mathfrak{g}/\mathfrak{h}}, \hspace{1 cm} v,w \in T_{x}M$$
where $\tilde{\theta}_{p}(x):T_{x}M \rightarrow \mathfrak{g}/\mathfrak{h}$. The
$H$-invariance of $\left\langle\cdot,
\cdot\right\rangle_{\mathfrak{g}/\mathfrak{h}}$ implies that $g^{\theta}(v,w)$
does not depend on the chosen frame $p$ over $x$. In this way, the translational part
$\theta$ of the Cartan connection $A$, by inducing an isomorphism between $TM$ and $P_{H} \times_{H}\mathfrak{g}/\mathfrak{h}$, permit us to define a metric $g^{\theta}$ on $M$ by using the $ad(H)$-invariant metric on $\mathfrak{g}/\mathfrak{h}$. To sum up, we can say that the metric $g^{\theta}$ that defines the
fundamental variable of the standard metric formulation of general relativity
is encoded in the $\mathfrak{g}/\mathfrak{h}$-component of the Cartan gauge
field $A$.

It is worth reminding that the frame bundle $\mathcal{F}M$ is itself endowed
with a (horizontal and $GL(\mathfrak{g}/\mathfrak{h})$-equivariant)
\emph{canonical} (or \emph{tautological}) \emph{form} $\theta_{c}\in
\Omega^{1}(\mathcal{F}M, \mathfrak{g}/\mathfrak{h})$. This form is defined by
means of the following expression (\cite{kobayashi-nomizu}, p.119):
\begin{align*}
\theta_{c}: T_{e(x)}(\mathcal{F}M) &\rightarrow \mathfrak{g}/\mathfrak{h}
\\ v &\mapsto e(x)^{-1}(\pi_{\ast}(v)),
\end{align*}
where $e(x):\mathfrak{g}/\mathfrak{h}\rightarrow T_{x}M$ is a frame on
$T_{x}M$. The definition of $\theta_{c}$ depends in an essential manner on the
fact that the fibers of $\mathcal{F}M$ are composed of frames on the tangent
spaces to $M$. This means that the existence of $\theta_{c}$ simply attests the
fact that the bundles $\mathcal{F}M$ and $TM$, far from being arbitrary bundles
on $M$, are \emph{naturally} related to the geometry of $M$ itself. By using
the isomorphism (\ref{ynp}), we can define a \emph{geometric canonical form}
$\tilde{\theta}_{c}: TM \rightarrow \mathcal{F}M
\times_{GL(\mathfrak{g}/\mathfrak{h})}\mathfrak{g}/\mathfrak{h}\simeq
TM$. It is easy to see that
$\tilde{\theta}_{c}$ is nothing but the identity in $TM$.\footnote{Indeed,
$\tilde{\theta}(\tilde{v})=q(e,\theta_{e}(v))=q(e,e^{-1}(\pi_{\ast}v))=q(e,e^{-1}\tilde{v})=
q(e, (\tilde{v}_{1},...,\tilde{v}_{n}))=v$, where $v\in T_{e}(\mathcal{F}M)$ is
any lift of $\tilde{v}$ and where we have used expression (\ref{icy}) in
Appendix \ref{appII}.} In this way, the canonicity of $\theta_{c}$ is reflected
by the fact that it merely defines a tautological identification of $TM$ with
itself. In the literature, the terms \emph{soldering form} and\emph{ canonical
form} are sometimes used as exchangeable terms. However, the previous discussion clearly shows that these two terms must be carefully
distinguished. In fact, it can be shown that the soldering form $\theta \in
\Omega^{1}(P_{H}, \mathfrak{g}/\mathfrak{h})$ is the pullback by $f^{\theta}$
of the canonical form $\theta_{c}$ on $\mathcal{F}M$ \cite{sternberg-1985}. The
important point that we want to stress here is that, contrary to the canonical
$1$-form $\theta_{c}$ on $\mathcal{F}M$, the soldering form $\theta$ on $P_{H}$
is not canonically defined. Indeed, $\theta$ is the
$\mathfrak{g}/\mathfrak{h}$-valued component of a Cartan
connection on $P_{H}$ that is not canonically defined.\footnote{This point was also stressed in
Ref.\cite{petti} (p.742). See also Ref.\cite{sharpe} (footnote in p.363) and
Ref.\cite{witten-1988} (p.51). It is also worth
noting that the non-canonicity of $\theta$ implies that we cannot
establish a bijective correspondence between the Ehresmann
connections $\omega_{G}$ on the bundle $P_{G}$ of \emph{affine} frames and the Ehresmann connections
$\omega_{H}$ on the bundle $P_{H}$ of \emph{linear} frames (like the bijective correspondence established in Ref.\cite{kobayashi-nomizu}, Theorem 3.3, p.129).} This is
consistent with the fact that in gravitational theories $\theta$ is assumed to
define a degree of freedom of the theory inducing a particular metric on $M$.
On the contrary, a geometric structure that is canonically defined (such as
$\theta_{c}$) is a fixed structure that cannot define dynamical degrees of
freedom.

\section{On the relation between external diffeomorphisms and internal gauge translations}
\label{sec: diffeos}

The difference between the group of \emph{external} diffeomorphisms of
spacetime and the gauge group of \emph{internal} local gauge transformations of a
Yang-Mills theory is an important obstruction to the comprehension of the
gravitational interaction in terms of a gauge theory. Whereas the gauge group
acts internally on the fibers at each $x\in M$, the group $Diff(M)$  acts
externally in the sense that it transforms spatiotemporal locations into one
another. We shall now explain in what sense the translational component of the
Cartan connection (i.e. the soldering form) permits us to establish a link
between \emph{external} diffeomorphisms of $M$ and \emph{internal} translations
in the different LHM.

The $\mathfrak{h}$-valued component of the Cartan connection $A$, i.e. the
Ehresmann connection $\omega_{H}$ on $P_{H}$, defines parallel transports in the
bundle $$P_{H} \times_{H}\mathfrak{g}/\mathfrak{h} \simeq V_{\sigma}(P_{G}\times _{G} G/H)$$ associated to $P_{H}$. In other terms, $\omega_{H}$ defines parallel transports of vectors tangent to the LHM $M_{0}^{x}$ along $\sigma$. Since the form $\tilde{\theta}$ defines an
isomorphism $TM\xrightarrow{\simeq} P_{H} \times_{H}\mathfrak{g}/\mathfrak{h}$, the Ehresmann connection $\omega_{H}$
transports vectors tangent to $M$ as expected. Now, while the $\mathfrak{h}$-valued
component $\omega_{H}$ of $A$ defines parallel transports of ``internal''
elements (tangent vectors to $M$ in the present case) as in Yang-Mills theory, the
$\mathfrak{g}/\mathfrak{h}$-valued component $\theta$ ``transports'' (as we shall now explain) the
spatiotemporal locations themselves. In order to show that this is the case, we shall now introduce the Cartan's notion of \emph{development}.

Let's consider a curve $\gamma: [0,1] \rightarrow M$ starting at $x_{0}\in M$ and let $\tilde{\gamma}: [0,1] \rightarrow P_{H}$ be any lift of $\gamma$ to $P_{H}\xrightarrow{\pi} M$
(that is, $\pi(\tilde{\gamma}(t))=\gamma(t)$ for all $t\in [0,1]$). 
Since $P_{H}\subset  P_{G}$, the curve $\tilde{\gamma}$ is included in
$P_{G}\xrightarrow{\pi'} M$. If we use the Ehresmann connection $\omega_{G}$ on
$P_{G}$ for parallel transporting $\tilde{\gamma}(t)$ to $\pi'^{-1}(x_{0})$
along $\gamma$ for all $t\in [0,1]$, we obtain a curve $\hat{\gamma}$ in
$\pi'^{-1}(x_{0})$. By using the projection $P_{G}\xrightarrow{\varrho} P_{G}/H
\simeq P_{G}\times_{G} G/H$, we can define a curve
$\gamma^{\ast}=\varrho(\hat{\gamma})$ in the fiber of $P_{G}\times_{G} G/H$
over $x_{0}$ called the \emph{development of $\gamma$ over
$x_{0}$}.\footnote{It can be shown that $\gamma^{\ast}$ only depends on
$\gamma$ and that it is independent from the choice of the lift
$\tilde{\gamma}$ (\cite{kobayashi-connections}, $\S$5).} Since the development
of a curve is obtained by projecting a parallel transport defined by
$\omega_{G}$ onto fibers isomorphic to $G/H$, the notion of development only
depends on the $\mathfrak{g}/\mathfrak{h}$-valued part of $\omega_{G}$, that is
$\theta$. The development process maps curves $\gamma$ in $M$ starting at
$\gamma(0)$ into curves $\gamma^{\ast}$ in the internal LHM $M_{0}^{x_{0}}$
over $\gamma(0)$. This process can be understood as the result of ``rolling''
back the LHM $M_{0}^{\gamma(1)}$ along $\gamma$ until $\gamma(0)$. This means
that the point $\gamma^{\ast}(1)$ in $M_{0}^{x_{0}}$ is the position reached by
the point of attachment of the LHM $M_{0}^{\gamma(1)}$ when the latter is
rolled back to $\gamma(0)$ along $\gamma$.\footnote{It is worth noting that the
condition (\ref{xus}) guarantees that the lifted curved $\tilde{\gamma}$ cannot
be a horizontal lift with respect to $\omega_{G}$. If $\tilde{\gamma}$ were a
$\omega_{G}$-horizontal lift of $\gamma$, the parallel transport of the points
in $\tilde{\gamma}(t)$ to $\pi'^{-1}(x_{0})$ along $\gamma$ would just
transport them back to $\tilde{\gamma}(0)$. Instead of obtaining a non-trivial
curve in $M_{0}^{x_{0}}$, the development of $\gamma$ would just be the
constant curve $\gamma^{\ast}(t)=\sigma(x_{0})$ for all $t\in [0,1]$. This
means that such a ``development'' would identify the point of attachment at
$\gamma(1)$ with the point of attachment at $\gamma(0)=x_{0}$. Now, this
amounts to translate the LHM $M_{0}^{\gamma(1)}$ back to $x_{0}$ by
``slipping'' it along $\gamma$. Roughly speaking, the condition (\ref{xus})
guarantees that the notion of Cartan's development encodes the idea of
``rolling without slipping'' the local Klein models along curves in $M$.}  In
this way, the translational part $\theta$ of $A$ defines the $\gamma$-dependent
image of any $x\in M$ in the LHM at $x_{0}$. In imaged terms, we could say that
each LHM $M_{0}^{x}$ is a sort of internal ``monad'' placed at $x$ wherein we
can ``print'' an image of every path in $M$ starting at $x$.

Let's consider the notion of development from an infinitesimal viewpoint. Given
a field of \emph{external} displacements $\tilde{v}\in TM$, the form
$\tilde{\theta}:TM\rightarrow E\simeq V_{\sigma}(P_{G}\times _{G} G/H)$ defines
a field of \emph{internal} displacements in the different LHM $M_{0}^{x}$ at
the points of attachment defined by $\sigma$, that is
$\tilde{\theta}(\tilde{v}(x))\in T_{\sigma(x)}M_{0}^{x}$. This means that the
point of attachment of the LHM $M_{0}^{x+\tilde{v}(x)}$ will be developed into
the point $\sigma(x)+ \tilde{\theta}(\tilde{v}(x))$ of the LHM $M_{0}^{x}$ when
$M_{0}^{x+\tilde{v}(x)}$ is rolled back to $x$ along the displacement $v(x)$
(\cite{stelle-west-1980}, Section IV). In other terms, the point
$\sigma(x)+\tilde{\theta}(\tilde{v}(x))$ ``represents'' the attaching point
$x+\tilde{v}(x)$ of the LHM $M_{0}^{x+\tilde{v}(x)}$ in the LHM $M_{0}^{x}$.
Different soldering forms identify $\sigma(x)+\tilde{\theta}(\tilde{v}(x))$
with different points in $M_{0}^{x}$. In this way, the soldering form
$\tilde{\theta}$ defines a correspondence between \emph{external spatiotemporal
diffeomorphisms} generated by vector fields $\tilde{v}$ on $M$ and
\emph{internal} \emph{gauge translations} in the LHM generated by
$\tilde{\theta}(\tilde{v})\in \Gamma(E)$. In other terms, the soldering form
$\tilde{\theta}$ permit us to ``internalize'' the diffeomorphisms of $M$ (\cite{Gronwald-1997}, Section 2.4). Reciprocally, the soldering form permit
us to interpret a field of internal gauge translations $\upsilon \in \Gamma(E)$
(where $\upsilon(x)\in T_{\sigma(x)}M_{0}^{x}$) as a generator $\tilde{\theta}^{-1}(\upsilon)$ of an external infinitesimal
diffeomorphism of $M$. Now, the field $\upsilon
\in \Gamma(E)$ can be interpreted as an infinitesimal transformation of the
attaching section $\sigma$. In turn, the point $\sigma(x)+\upsilon(x)$ in the LHM
$M_{0}^{x}$ becomes an attaching point when the Klein geometry $(M_{0}^{x},
\sigma(x))$ is rolled forward along the external infinitesimal displacement
$\tilde{\theta}^{-1}(\upsilon)$. Therefore, the soldering process allows us to interpret
the transformed section $\sigma+\upsilon$ as the attaching section of the geometry obtained by rolling forward along $\tilde{\theta}^{-1}(\upsilon)$ all the LHM. Briefly, \emph{we can
change the attaching section by rolling the LHM along diffeomorphisms of $M$} (a similar argument was proposed in Ref.\cite{Gronwald-1998}, Section 4).
Therefore, the invariance of a theory under $Diff(M)$ guarantees its invariance
under transformations of the attaching section $\sigma$, i.e. under changes of
the partial gauge fixing that defines the attachment of the local models to
$M$.

\section{The Adjoint Tractor Bundle}
\label{sec: tractor}

In the last section, we have obtained a gauge-theoretical description of the
group $Diff(M)$, that is we have shown that an infinitesimal diffeomorphism of
$M$ can be interpreted as an internal local gauge transformation defined by a
section of the vector bundle
$P_{H}\times_{H}\mathfrak{g}/\mathfrak{h}\rightarrow M$. We shall now consider
this result in the framework provided by the Atiyah algebroid associated to the
bundle $P_{H}\rightarrow M$. As we have shown in Section N$^{\circ}$\ref{sect:
ehresmann}, the vertical automorphisms of $P_{H}$ (or local gauge
transformations) are generated by the sections of the Lie algebra bundle
$P_{H}\times_{H}\mathfrak{h}$, that is $aut_{v}(P_{H})=\Gamma(P_{H}\times_{H}\mathfrak{h})$. In turn, the general automorphisms of $P_{H}$
are generated by the $H$-invariant vector fields on $P_{H}$, that is $aut(P_{H})=\Gamma^{H}(TP_{H})$. By using
(\ref{nfe})) we can also express $aut(P_{H})$ in terms of the sections of the Atiyah algebroid $TP_{H}/H$, that is $aut(P_{H})=\Gamma(TP_{H}/H)$. Whereas $aut_{v}(P_{H})$ is naturally expressed in terms of the sections of a Lie algebra bundle (with fibers modeled on $\mathfrak{h}$), this is not the case for $aut(P_{H})$. In other terms, $aut(P_{H})$ cannot be \emph{naturally} expressed in gauge-theoretical terms, i.e. as an action obtained by
localizing a Lie group action. Now, we can provide such a gauge-theoretical description of
$Aut(P_{H})$ if $P_{H}$ is equipped with a
Cartan connection. Indeed, a Cartan connection $A$ defines an isomorphism of vector
bundles between the Atiyah algebroid $TP_{H}/H$ and the so-called \emph{adjoint
tractor bundle} $P_{H}\times_{H}\mathfrak{g}\rightarrow M$
(\cite{Crampin-2009}, Theorem 1). The isomorphism $A:TP_{H}\rightarrow
\mathfrak{g}$ defined by the Cartan connection induces an $H$-equivariant isomorphism
$\Theta: TP_{H}\rightarrow P_{H}\times \mathfrak{g}$ of vector bundles over the identity on $P_{H}$ given by $X\in T_{p}P_{H}\mapsto (p, A_{p}(X))$. This isomorphism is $H$-equivariant in the sense that $\Theta((R_{g})_{\ast}X)=(pg, A_{pg}((R_{g})_{\ast}X))=(pg, Ad_{g^{-1}} A_{p}(X))=(p, A_{p}(X))\cdot g$. Hence, we can take the quotient by the action of $H$, thereby obtaining the following  isomorphism
\begin{eqnarray}\label{ber}
TP_{H}/H \simeq P_{H}\times_{H} \mathfrak{g}
\end{eqnarray}
of bundles over the identity on $P_{H}/H \simeq M$ (\cite{Mackenzie-2005}, Proposition 3.1.2(ii), p.88). Thanks to this isomorphism the adjoint tractor bundle $P_{H}\times_{H} \mathfrak{g}$ acquires the structure of a (transitive) Lie algebroid\footnote{Firstly, the bracket of vector fields on $P_{H}$ induces a
bracket $[\![ \cdot , \cdot ]\!]$ in $\Gamma(P_{H}\times_{H}\mathfrak{g})$ by
means of the expression $[\![ \sigma_{A(X)},\sigma_{A(Y)} ]\!]=\sigma_{A([X,Y])}$. Secondly, the anchor $\Pi:P_{H}\times_{H}\mathfrak{g} \rightarrow TM$ is given by $[(p,\xi)]_{H} \mapsto \pi_{\ast}(A_{p}^{-1}(\xi))$, being its kernel the subbundle $P_{H}\times_{H}\mathfrak{h}$.}, and the Atiyah algebroid $TP_{H}/H$ acquires the structure of a Lie algebra bundle (\cite{Crampin-2009}, Section 5). In this way, the Cartan connection $A$ permits us to describe the infinitesimal automorphism
of $P_{H}$ (originally given by the sections of $TP_{H}/H$) in terms of the
sections of the Lie algebra bundle $P_{H}\times_{H} \mathfrak{g}$.

A section of
the adjoint tractor bundle $P_{H}\times_{H}\mathfrak{g}$ can be directly
obtained from an $H$-invariant vector field in $P_{H}$ as follows. Given an
$H$-invariant vector field $X\in \Gamma^{H}(TP_{H})$ generating an
infinitesimal automorphism of $P_{H}$, the Cartan connection $A$ defines a
$\mathfrak{g}$-valued $H$-equivariant function $A(X) \in\mathcal{C}^{H}(P_{H},
\mathfrak{g})$. In turn, these equivariant functions are in bijective
correspondence with the sections of the adjoint tractor bundle
$P_{H}\times_{H}\mathfrak{g}\rightarrow M$ (see Appendix
N$^{\circ}$\ref{appI}). In this way, the Cartan connection induces a bijection
\begin{eqnarray*}
aut(P_{H})=\Gamma^{H}(TP_{H}) &\xrightarrow{\simeq}& \Gamma(P_{H}\times_{H}\mathfrak{g})
\\ X &\mapsto& \sigma_{A(X)},
\end{eqnarray*}
between the infinitesimal generators of the automorphisms of $P_{H}$
and the sections of a Lie algebra bundle.

We can summarize
the relations between the different bundles by saying that the Cartan
connection $A=\omega_{H}+\theta$ induces the following isomorphism of exact sequences of vector
bundles:

$$
  \xymatrix@R=.5in @C=.8in{
      & & P_{H}\times_{H}\mathfrak{g} \ar@{->>}[r] & P_{H}\times_{H}\mathfrak{g}/ \mathfrak{h} \ar[rd] &  \\
    0 \ar[r] &  P_{H}\times_{H}\mathfrak{h} \ar@{^{(}->}[ru] \ar@{^{(}->}[rd] & &  & 0, \\
      &   & TP_{H}/H \ar@{.>}[uu]^{A} \ar@{->>}[r] & TM \ar@/^1.2pc/[l]^{\gamma_{\omega_{H}}} \ar[ru] \ar@{.>}[uu]^{\tilde{\theta}} &
  }
$$
where $\gamma_{\omega_{H}}$ is the Lie algebroid connection associated to the
Ehresmann connection $\omega_{H}$. The bottom exact sequence is canonically associated to the principal bundle $P_{H}\rightarrow M$. It encodes the
relations between the vertical automorphisms of $P_{H}$ (generated by elements in $aut_{v}(P_{H})=\Gamma(P_{H}\times_{H}\mathfrak{h})$), the
automorphisms of $P_{H}$ (generated by elements in $aut(P_{H})=\Gamma(TP_{H}/H)$), and the diffeomorphisms of $M$. If $P_{H}$ is in addition endowed with a Cartan connection, then we can describe $aut(P_{H})$ and $\Gamma(TM)$ in terms of the sections of the Lie algebra bundle $P_{H}\times_{H}\mathfrak{g}$ and the vector bundle $P_{H}\times_{H}\mathfrak{g}/ \mathfrak{h}$ respectively.

\section{Transformations of the fields under Local Lorentz transformations and local gauge translations}
\label{sec: transformations}

Since $P_{H}$ is endowed with an Ehresmann connection $\omega_{H}$, we can
split an infinitesimal automorphism of $P_{H}$ into a vertical automorphism and
(the horizontal lift defined by $\omega_{H}$ of) an infinitesimal
diffeomorphism of $M$. In other terms, the $H$-invariant vector fields
generating infinitesimal automorphisms of $P_{H}$ can be decomposed as
$v=v_{h}+v_{v}$ where $v_{h}$ is the horizontal lift of $\tilde{v}\in TM$ (i.e.
$\omega_{H}(v_{h})=0$) and $v_{v}\in VP_{H}$ is a vertical fundamental vector
field defined by an element in $\mathfrak{h}$ (by means of the map $\kappa:
\mathfrak{h} \rightarrow VP_{H}$). The vertical component $v_{v}$ defines local
gauge transformations (or vertical automorphisms of $P_{H}$) given by the
$H$-equivariant function $$\Lambda(p)=\omega_{H} (v_{v}(p))\in
\mathcal{C}^{H}(P_{H}, \mathfrak{h})$$ or, equivalently, by the corresponding
section in $\Gamma(P_{H}\times_{H}\mathfrak{h})$. Now, $P_{H}$ is not only
endowed with an Ehresmann connection $\omega_{H}$, but also with a (horizontal)
soldering form $\theta: TP_{H}\rightarrow \mathfrak{g}/\mathfrak{h}$. Hence, we
can understand the horizontal component $v_{h}$ (encoding an infinitesimal
diffeomorphism of $M$) in an analogous fashion. Indeed, the horizontal vector
field $v_{h}$ defines the $H$-equivariant function
$$\xi(p)=\theta_{p}(v_{h}(p))\in \mathcal{C}^{H}(P_{H},
\mathfrak{g}/\mathfrak{h}),$$ or, equivalently, the corresponding section in
$\Gamma(P_{H}\times_{H}\mathfrak{g}/\mathfrak{h})$.\footnote{It is worth noting
that we can directly define a section in
$\Gamma(P_{H}\times_{H}\mathfrak{g}/\mathfrak{h})$ directly from $\tilde{v}\in
TM$ by means of the geometric soldering form $\tilde{\theta}$.} While $\Lambda$
generates infinitesimal local Lorentz transformations (or vertical
automorphisms of $P_{H}$), the function $\xi$ encodes the infinitesimal change
of the attaching section $\sigma$ induced by the infinitesimal diffeomorphism
of $M$ generated by $\tilde{v}=\pi_{\ast}v_{h}\in TM$.

We shall now compute the transformations of the relevant fields under both
infinitesimal local Lorentz transformations and infinitesimal changes of the
attaching section (or local gauge translations). The transformations of the
fields $\theta$ and $\omega_{H}$ under an infinitesimal automorphism of $P_{H}$
generated by $v$ are given by the following Lie derivatives\footnote{Indeed,
\begin{eqnarray}\label{ixq}
\mathcal{L}_{v}A&=& i_{v}dA+d(A(v)) = i_{v}(d^{A}A-\frac{1}{2}[A,A])+d(\omega_{H}(v)+\theta(v)),
\\ &=& i_{v}F-\frac{1}{2}([A(v),A]-[A,A(v)])+d\Lambda+d\xi =i_{v}F+[A,A(v)]+d\Lambda+d\xi,\nonumber
\\ &=& i_{v}(R+T+\frac{1}{2}[\theta,\theta])+[\omega_{H}+\theta, \Lambda+\xi]+d\Lambda+d\xi,\nonumber
\\ &=& i_{v}R+i_{v}T-[\theta, \xi]+ [\omega_{H}, \Lambda]+[\omega_{H}, \xi]+ [\theta, \Lambda]+ [\theta, \xi]+ d\Lambda+d\xi,\nonumber
\\ &=& \underbrace{i_{v}R+d^{\omega_{H}}\Lambda}_{\in \Omega^{1}(P_{H}, \mathfrak{h})}+\underbrace{i_{v}T+ [\theta, \Lambda]+d^{\omega_{H}}\xi}_{\in \Omega^{1}(P_{H}, \mathfrak{g}/\mathfrak{h})}.\nonumber
\end{eqnarray}}:

\begin{align}\label{ixe}
\tilde{\delta}_{(\Lambda, \xi)}\theta \doteq \mathcal{L}_{v}\theta&= i_{v}T + [\theta, \Lambda]+ d^{\omega_{H}}\xi,
\\ \tilde{\delta}_{(\Lambda, \xi)}\omega_{H} \doteq \mathcal{L}_{v}\omega_{H}&= i_{v}R + d^{\omega_{H}}\Lambda.\nonumber
\end{align}

If the automorphism is purely vertical (i.e. a local Lorentz transformation
generated by a fundamental vector field $v=v_{v}$), then we have
\begin{eqnarray}\label{bfd}
\tilde{\delta}_{\Lambda}\theta &=& [\theta, \Lambda],
\\ \tilde{\delta}_{\Lambda}\omega_{H}&=& d^{\omega_{H}}\Lambda \nonumber
\end{eqnarray}
where we have used that the torsion $T$ and the curvature $R$ are horizontal
forms. In turn, the corresponding local Lorentz transformations of the
curvature and the torsion are given by\footnote{Indeed,
\begin{eqnarray*}\label{uxs}
\mathcal{L}_{v_{v}}R &=& i_{v_{v}}dR + d(R(v_{v})) = i_{v_{v}}d^{\omega_{H}}R -[\omega_{H}(v_{v}), R]+[\omega_{H}, R(v_{v})]= [R, \Lambda],\nonumber
\end{eqnarray*}
and
\begin{eqnarray*}\label{akt}
\mathcal{L}_{v_{v}}T &=& i_{v_{v}}dT + d(T(v_{v})) = i_{v_{v}}d^{\omega_{H}}T -[\omega_{H}(v_{v}), T]+[\omega_{H}, T(v_{v})],\nonumber
\\ &=& i_{v_{v}}d^{\omega_{H}}T +[T, \Lambda] = i_{v_{v}}[R,\theta] +[T, \Lambda],\nonumber
\\ &=& [R(v_{v}),\theta] - [R,\theta(v_{v})]+[T, \Lambda] =  [T, \Lambda].
\end{eqnarray*}}:
\begin{eqnarray*}\label{nde}
\tilde{\delta}_{\Lambda}R &=& [R, \Lambda],\nonumber
\\ \tilde{\delta}_{\Lambda}T &=& [T, \Lambda],
\end{eqnarray*}

If the automorphism is purely horizontal (i.e. $v=v_{h}$), then the
transformations of the gauge fields are
\begin{eqnarray}\label{vfr}
\tilde{\delta}_{\xi}\theta &=& i_{v_{h}}T + d^{\omega_{H}}\xi.
\\\tilde{\delta}_{\xi}\omega_{H}&=& i_{v_{h}}R \nonumber
\end{eqnarray}
and the transformations of the curvature and the torsion are\footnote{Indeed,
\begin{eqnarray*}\label{vzr}
\mathcal{L}_{v_{h}}R &=& i_{v_{h}}dR + d(R(v_{h}))= i_{v_{h}}d^{\omega_{H}}R -[\omega_{H}(v_{h}), R]+[\omega_{H}, R(v_{h})]+d(R(v_{h})),
\\ &=& d^{\omega_{H}}(R(v_{h})),\nonumber
\\ \mathcal{L}_{v_{h}}T &=& i_{v_{h}}dT + d(T(v_{h}))= i_{v_{h}}d^{\omega_{H}}T -[\omega_{H}(v_{h}), T]+[\omega_{H}, T(v_{h})]+d(T(v_{h})),\nonumber
\\ &=& i_{v_{h}}[R,\theta]+d^{\omega_{H}}T(v_{h})= [R(v_{h}),\theta]-[R,\xi] +d^{\omega_{H}}T(v_{h}),
\end{eqnarray*}}
\begin{eqnarray}\label{vzr}
\tilde{\delta}_{\xi}R &=& d^{\omega_{H}}(R(v_{h})),
\\ \tilde{\delta}_{\xi}T &=& [R(v_{h}),\theta]-[R,\xi] +d^{\omega_{H}}T(v_{h}),\nonumber
\end{eqnarray}
respectively.

We shall now compare the transformations (\ref{ixe}) of the gauge fields
$\theta$ and $\omega_{H}$ with the transformations obtained by performing local
gauge transformations of the bundle $P_{G}$. Indeed, an alternative strategy for studying the transformations of the spin connection $\omega_{H}$ and the soldering form $\theta$ is to simply restrict the local gauge transformations of $\omega_{G}\in \Omega^{1}(P_{G}, \mathfrak{g})$ to $P_{H}$ (this is the strategy followed in Ref.\cite{egeileh}, Section 5.2, and in Ref\cite{witten-1988}). As we have explained in Section
N$^{\circ}$\ref{sect: ehresmann}, the natural representation of a local gauge
transformation $\lambda\in \mathcal{C}^{G}(P_{G}, \mathfrak{g})$ on an
Ehresmann connexion $\omega_{G}$ in $P_{G}$ is given by expression (\ref{yrd}),
that is $\delta_{\lambda}\omega_{G}=d^{\omega_{G}}\lambda$. By taking the
pullback of this expression by the inclusion $\iota: P_{H}\hookrightarrow
P_{G}$ we obtain the expression
\begin{eqnarray}\label{fjr}
\delta_{\hat{\lambda}}A=d\hat{\lambda}+[A,\hat{\lambda}],
\end{eqnarray}
where $A\doteq \iota^{\ast}\omega_{G}=\omega_{G}|_{P_{H}}$ and
$\hat{\lambda}=\iota^{\ast}\lambda=\lambda|_{P_{H}}\in \mathcal{C}^{H}(P_{H},
\mathfrak{g})$. By using the decomposition $\mathfrak{g}=\mathfrak{h}\oplus
\mathfrak{g}/\mathfrak{h}$, we can define the following $H$-equivariant
functions on $P_{H}$
\begin{eqnarray*}\label{bde}
\Lambda &\doteq& \pi_{\mathfrak{h}}\circ \hat{\lambda} \in \mathcal{C}^{H}(P_{H},
\mathfrak{h})
\\ \xi &\doteq& \pi_{\mathfrak{g}/\mathfrak{h}}\circ \hat{\lambda} \in
\mathcal{C}^{H}(P_{H}, \mathfrak{g}/\mathfrak{h}).
\end{eqnarray*}

By substituting $A=\omega_{H}+\theta$ and $\hat{\lambda}=\Lambda+\xi $ in
(\ref{fjr}) we obtain the expressions (\cite{egeileh}, Section 5.2)\footnote{Indeed,
\begin{eqnarray*}
\delta_{\hat{\lambda}}A &=& d(\Lambda+\xi)+[\omega_{H}+\theta ,\Lambda+\xi]=d\Lambda+[\omega_{H},\Lambda]+ d\xi+[\omega_{H},\xi]+[\theta ,\Lambda]+[\theta ,\xi],
\\ &=& \underbrace{[\theta ,\xi]_{\mathfrak{h}}+ d^{\omega_{H}}\Lambda}_{\in \Omega^{1}(P_{H}, \mathfrak{h})}+\underbrace{[\theta ,\xi]_{\mathfrak{g}/\mathfrak{h}}+[\theta ,\Lambda]+d^{\omega_{H}}\xi}_{\in \Omega^{1}(P_{H}, \mathfrak{g}/\mathfrak{h})}.
\end{eqnarray*}}:
\begin{align}\label{beu}
\delta_{(\Lambda, \xi)} \theta &= [\theta, \xi]_{\mathfrak{g}/\mathfrak{h}} +[\theta, \Lambda] + d^{\omega_{H}}\xi,
\\ \delta_{(\Lambda, \xi)} \omega_{H} &= [\theta, \xi]_{\mathfrak{h}} + d^{\omega_{H}}\Lambda. \nonumber
\end{align}

The first terms on the right can be written as $[\theta,
\xi]_{\mathfrak{g}/\mathfrak{h}}=i_{v}T_{0}$ and $[\theta,
\xi]_{\mathfrak{h}}=i_{v}R_{0}$ respectively, where we have used the
expressions (\ref{xor}) for the curvature and the torsion of a flat Cartan
geometry. If the transformation is a pure Lorentz rotation (i.e. $\xi=0$), we
reobtain expressions (\ref{bfd}). If the transformation is a pure translation
(i.e. $\Lambda=0$), we have
\begin{eqnarray}\label{dze}
\delta_{\xi} \theta &=& i_{v}T_{0} + d^{\omega_{H}}\xi,
\\ \delta_{\xi} \omega_{H} &=& i_{v}R_{0}. \nonumber
\end{eqnarray}

These transformations have to be compared with (\ref{vfr}). Since the
transformations (\ref{dze}) were obtained by restricting the \emph{vertical} local gauge
transformations of the bundle $P_{G}$ to $P_{H}$, it is natural that they depend on the
curvature $R_{0}$ and the torsion $T_{0}$ of the Cartan-flat LHM. On the
contrary, the transformations (\ref{vfr}) of $(\omega_{H}, \theta)$ under local
gauge translations depend on the curvature $R$ and the torsion $T$ of $M$
itself. This is natural, since these local gauge translations are identified,
via the soldering procedure, with the diffeomorphisms of $M$. It is worth comparing the computation (\ref{yrd}) of the local gauge transformations of an Ehresmann connection with the computation (\ref{ixq}). In (\ref{yrd}), the curvature terms vanishes since the gauge transformations are purely vertical. On the contrary, the terms $i_{v}R$ and $i_{v}T$ do not vanish in (\ref{ixq}), since  the vector field $v$, far from being purely vertical, has an horizontal component $v_{h}$ related to a diffeomorphism of $M$. Both kinds of
transformations coincide when the Cartan geometry $(P_{H}, A)$ is Cartan-flat.
Indeed, the difference between (\ref{dze}) and (\ref{vfr}) is given by the
components of the Cartan curvature:
\begin{align}\label{bzl}
\tilde{\delta}_{\xi}\theta - \delta_{\xi} \theta  &= i_{v}T - [\theta, \xi]_{\mathfrak{g}/\mathfrak{h}}=i_{v}F_{\mathfrak{g}/\mathfrak{h}}
\\ \tilde{\delta}_{\xi}\omega_{H}-\delta_{\xi} \omega_{H} &= i_{v}R - [\theta, \xi]_{\mathfrak{h}} =i_{v}F_{\mathfrak{h}}.\nonumber
\end{align}

Therefore, we can indistinctly use both kinds of transformations when $(P_{H},A)$ is Cartan flat. However, the invariance under local gauge translations will be the encoded by the transformations (\ref{vfr})--rather than (\ref{dze})--in the general case.

\section{Conclusion}

In order to conclude, we shall recapitulate the construction discussed in this paper. We have started with a $G$-principal bundle $P_{G}\rightarrow M$ endowed with an Ehresmann connection $\omega_{G}$. A fiber in this bundle is isomorphic to the set of \emph{affine frames} $(m,e)$ in the Klein geometries associated to the pair $(G,H)$ (where an affine frame is given by a point $m$ in the Klein geometry and a linear frame $e$ of the tangent space to $m$). In order to
model $M$ by means of tangent Klein geometries associated to the pair $(G,H)$
we must \emph{attach} (zero-order identification) and \emph{solder} (first-order identification) copies of $G/H$ to $M$. To do so, we
have firstly defined a bundle $P_{G}\times_{G}G/H \rightarrow M$ associated to $P_{G}$ with fibers
$M_{0}^{x}$ isomorphic to $G/H$. We have then defined an \emph{attaching section}
$\sigma: M\rightarrow P_{G}\times_{G}G/H$ that selects an attaching point
$\sigma(x)$ in each fiber $M_{0}^{x}$. By doing so, we have obtained a bundle of
Klein geometries $(M_{0}^{x}, \sigma(x))$ attached to $M$ along $\sigma$. This attaching procedure induces a reduction of the $G$-principal bundle $P_{G}$ of \emph{affine frames} down to an $H$-principal bundle $P_{H}$ of \emph{linear frames}. In fact, the selection of a point $\sigma(x)$ in each homogeneous model $M_{0}^{x}$ breaks the affine symmetries of $M_{0}^{x}$ down to the Lorentz symmetry of the linear frames at $\sigma(x)$. The restriction of the Ehresmann connection $\omega_{G}$ to the reduced bundle $P_{H}\subset P_{G}$ defines a Cartan connection $A\doteq \omega_{G}|_{P_{H}}\in\Omega^{1}(P_{H}, \mathfrak{g})$ on $P_{H}$ (provided that the condition (\ref{xus}) is satisfied). The Cartan connection $A$ can be decomposed in an Ehresmann connection $\omega_{H}$ gauging the local Lorentz symmetry of $P_{H}$ (the so-called \emph{spin connection}) and a soldering form $\theta$. The later defines an identification between each tangent space $T_{x}M$ and the tangent space to the local homogeneous model $M_{0}^{x}$ at the attaching point $\sigma(x)$ (first-order identification between $M$ and $M_{0}^{x}$). This identification can be used to transfer the Lorentz-invariant scalar product in $\mathfrak{g}/\mathfrak{h}$ to $TM$, thereby defining a metric $g^{\theta}$ on $M$. The soldering induced by $\theta$ also permits to \emph{develop} any ``external'' path $\gamma \subset M$ starting at $x$ in the ``internal'' Klein geometry $(M_{0}^{x}, \sigma(x))$ at $x$. Infinitesimally, any vector field on $M$ can be ``lifted'' to a section of the vector bundle $P_{H}\times_{H}\mathfrak{g}/\mathfrak{h}$ encoding the vector spaces tangent to each $M_{0}^{x}$ at $\sigma(x)$. We have argued that a section of $P_{H}\times_{H}\mathfrak{g}/\mathfrak{h}$ can be interpreted as an infinitesimal transformation of the attaching section $\sigma$. In this way, an infinitesimal diffeomorphism of $M$ can be interpreted as an infinitesimal transformation of the ``partial gauge fixing'' defined by $\sigma$. The importance of this remark relies on the fact that the local translational invariance (explicitly broken by the election of an attaching section $\sigma$) is \emph{implicitly} preserved by the invariance under transformations of the partial gauge fixing defined by $\sigma$, which in turn is guaranteed (thanks to the soldering procedure) by the invariance under $Diff(M)$.

\section{Appendixes}

\subsection{Bijection between sections of associated bundles and pseudo-tensorial functions} \label{appI}
Given a $G$-principal bundle $P\rightarrow M$  and a left action $G\times
S\rightarrow S$, we shall demonstrate the following bijection
\begin{align}\label{nag}
\Gamma(P\times_{G} S)\cong \mathcal{C}^{G}(P, S),
\end{align}
where $\mathcal{C}^{G}(P, S)$ denotes the set of $G$-equivariant $S$-valued
functions on $P$, i.e. the functions $\varphi:P\rightarrow S$ that satisfy
$\varphi(pg)=g^{-1}\varphi(p)$ (\cite{kolar-michor-slovak}, p.94). Given such a
function, the induced section $\sigma_{\varphi}:M\rightarrow P\times_{G} S$ is
given by $\sigma_{\varphi}(x)=[(p, \varphi (p))]$, where $\pi(p)=x$. This
definition does not depend on the chosen $p$. Indeed, if we choose $p'=pg\in
\pi^{-1}(x)$, then $\sigma_{\varphi}(x)=[(pg, \varphi (pg))]=[(pg,
g^{-1}\varphi (p))]=[(p, \varphi (p))]$, where we have used the
$G$-equivariance of $\varphi$. Conversely, given a section $\sigma:M\rightarrow
P\times_{G} S$, we can define the following $G$-equivariant function:
\begin{align*}
\varphi_{\sigma}:P &\rightarrow S
\\ p &\mapsto \tau (p,\sigma(\pi(p))),
\end{align*}
where $\tau: P \times (P\times_{G} S) \rightarrow S$ is the map that sends the
frame $p$ and a geometric object in the associated bundle $P\times_{G} S$ to
the ``coordinates'' (in $S$) of this object in the frame $p$. The map $\varphi$
is indeed equivariant: $\varphi_{\sigma}(pg)=\tau (pg,\sigma(\pi(pg)))=\tau
(pg,\sigma(\pi(p)))=g^{-1}\varphi(p)$ (roughly speaking, if we rotate the frame
$p$ by $g$, the ``coordinates'' in $S$ change by $g^{-1}$). It can be shown
that the definition of $\sigma_{\varphi}$ from $\varphi$ and the definition of
$\varphi_{\sigma}$ from $\sigma$ are inverse of one another.

\subsection{Isomorphism between geometric and coordinate differential forms}
\label{appII} Given a tensorial $1$-form $\theta: TP_{H} \rightarrow
\mathfrak{g}/\mathfrak{h}$, the $1$-form $$\tilde{\theta}: TM \rightarrow E$$
is given by
\begin{align}\label{icy}
\tilde{\theta}(\tilde{v})=q(p, \theta_{p} (v))
\end{align}
where $v$ is any lift of $\tilde{v}$ to a point $p\in P_{H}$ such that
$\pi_{\ast}v=\tilde{v}$ and where $q: P_{H} \times
\mathfrak{g}/\mathfrak{h}\rightarrow E$ is the quotient map. It is easy to see
that this definition depends neither on the chosen frame $p\in \pi^{-1}x$ nor
on the chosen lift $v\in T_{p}P_{H}$. On the other hand, given a
$E$-valued $1$-form $\tilde{\theta}: TM \rightarrow E$ on $M$, the tensorial
form $\theta: TP_{H} \rightarrow \mathfrak{g}/\mathfrak{h}$ is given by
\begin{align}\label{xiz}
\theta_{p} (v)=\tau(p,\tilde{\theta}(\pi_{*}v)), \hspace{1 cm} v\in T_{p}P_{H}
\end{align}
with
\begin{align*}
\tau: P_{H} \times_{M} E &\rightarrow \mathfrak{g}/\mathfrak{h}
\\(p, \hat{v}) &\mapsto (\hat{v}_{1},...,\hat{v}_{n})
\end{align*} where $(\hat{v}_{1},...,\hat{v}_{n})$ are the
components of $\hat{v}\in E$ in the frame $p$. It can be shown that the $1$-form $\theta$
is horizontal (since $\pi_{*}v=0$ for vertical vectors) and $G$-equivariant.

\section*{Acknowledgments}

The research leading to these results has received funding from the European Research Council under the European Community's Seventh Framework Programme (FP7/2007-2013 Grant Agreement n$^{\circ}$ 263523).

\bibliographystyle{amsplain}
\providecommand{\bysame}{\leavevmode\hbox
to3em{\hrulefill}\thinspace}

\end{document}